\documentclass[%
aip,
superscriptaddress,
 amsmath,amssymb,
 reprint,%
 citeautoscript,
]{revtex4-2}

\usepackage{microtype}
\usepackage{graphicx}
\usepackage{caption}
\usepackage{subcaption}
\usepackage{booktabs} 
\usepackage{hyperref}
\usepackage{chemformula}
\usepackage{multirow} 

\usepackage{dcolumn}

\usepackage{amsmath}
\usepackage{amssymb}
\usepackage{mathtools}
\usepackage{amsthm}
\usepackage{bm}

\usepackage[capitalize,noabbrev]{cleveref}

\theoremstyle{plain}

\theoremstyle{definition}

\theoremstyle{remark}

\newcommand{\rv}{\mathbf{r}}
\newcommand{\Rv}{\mathbf{R}}
\newcommand{\xv}{\mathbf{x}}

\newcommand{\dd}{\mathrm{d}}

\newcommand{\E}{\mathbb{E}}

\DeclareMathOperator*{\argmin}{arg\,min}
\DeclarePairedDelimiter{\abs}{\lvert}{\rvert}

\newcommand{\methodabb}{NERD}

\begin{document}
\title{Highly Accurate Real-space Electron Densities with Neural Networks}
\author{Lixue Cheng}
\thanks{These authors contributed equally to this work.}
\affiliation{Microsoft Research, AI for Science}
\author{P. Bern\'{a}t Szab\'{o}}
\thanks{These authors contributed equally to this work.}
\affiliation{Microsoft Research, AI for Science}
\affiliation{Freie Universit\"{a}t Berlin}
\author{Zeno Sch\"{a}tzle}
\thanks{These authors contributed equally to this work.}
\affiliation{Microsoft Research, AI for Science}
\affiliation{Freie Universit\"{a}t Berlin}
\author{Derk P. Kooi}
\affiliation{Microsoft Research, AI for Science}
\author{Jonas K\"{o}hler}
\affiliation{Microsoft Research, AI for Science}
\author{Klaas J. H. Giesbertz}
\affiliation{Microsoft Research, AI for Science}
\author{Frank No\'{e}}
\email{franknoe@microsoft.com}
\affiliation{Microsoft Research, AI for Science}
\affiliation{Freie Universit\"{a}t Berlin}
\author{Jan Hermann}
\email{jan.hermann@microsoft.com}
\affiliation{Microsoft Research, AI for Science}
\author{Paola Gori-Giorgi}
\email{pgorigiorgi@microsoft.com}
\affiliation{Microsoft Research, AI for Science}
\author{Adam Foster}
\email{adam.e.foster@microsoft.com}
\affiliation{Microsoft Research, AI for Science}

\begin{abstract}
Variational ab-initio methods in quantum chemistry stand out among other methods in providing direct access to the wave function.
This allows in principle straightforward extraction of any other observable of interest, besides the energy, but in practice this extraction is often technically difficult and computationally impractical.
Here, we consider the electron density as a central observable in quantum chemistry and introduce a novel method to obtain accurate densities from real-space many-electron wave functions by representing the density with a neural network that captures known asymptotic properties and is trained from the wave function by score matching and noise-contrastive estimation.
We use variational quantum Monte Carlo with deep-learning ansätze (deep QMC) to obtain highly accurate wave functions free of basis set errors, and from them, using our novel method, correspondingly accurate electron densities, which we demonstrate by calculating dipole moments, nuclear forces, contact densities, and other density-based properties.
\end{abstract}
\maketitle

\section{Introduction}
\label{sec:intro}

Variational ab-initio methods form the bedrock of quantum chemistry, be it the Hartree--Fock method and variational quantum Monte Carlo (VMC) in first quantization or various flavors of configuration interaction (CI) in second quantization \citep{Piela14}.
They provide direct access to the wave function, unlike perturbation-based or projection-based methods.
Access to the full wave function in theory allows the extraction of any observable of interest, over and above the electronic energy, but calculating such observables is often computationally involved.
Consider for instance the extraction of real-space one- and two-electron densities from many-electron wave functions.
With real-space wave functions, such as those yielded by VMC, one has to carefully design low-variance estimators for the density, and then sample those estimators individually at each point at which the density is to be determined, limiting resolution and further use \citep{assaraf_improved_2007,VarBarGui-JCP-2014,hakansson_efficient_2008,per_zero-variance_2011}.
With second-quantized wave functions, such as those yielded by CI, one is often severely limited by the accuracy of the basis set.
The basis-set extrapolation techniques typically used for relative energies \citep{TruhlarCPL98,KartonJCC17} are not available for electron densities, and Gaussian basis sets have built-in incorrect asymptotic behavior at the nuclei as well as far away from them.
Non-variational methods (e.g.\ coupled cluster and  Møller–Plesset) are popular to improve the accuracy of the total energy, but extracting observables from them are non-trivial as there is no unique approach to do so. For example, in coupled cluster we have the (cheap) unrelaxed density matrices and the inequivalent relaxed density via the costly $\lambda$-equations \citep{KucharskiJCP98}.

In particular, the one-electron density $\rho(\rv)$ succinctly describes many electronic properties of matter, and it is a primary example of useful information that can be extracted from a many-electron wave function.
It is the key object in density functional theory, forms the basis of bond classification \citep{SilviN94}, and allows computation of other derived observables, such as electrostatic multipole moments, nuclear forces, or contact densities.
Here, we address the existing issues in obtaining accurate one-electron densities and introduce a novel method for their extraction from real-space many-electron wave functions.
The \emph{neural electron real-space density} (\methodabb) approach retains the accuracy of the underlying wave-function method; once fitted, it provides the density everywhere in space at negligible computational cost, and is asymptotically correct by design both at the nuclei and far away from them.
\methodabb\ represents the one-electron density as a neural network model with built-in physical constraints, and is fitted with a loss function composed of two terms on data consisting of spatial electron coordinates sampled from the target wave function.
The first loss term uses score matching \citep{hyvarinen2005estimation,vincent2011connection} to match gradients of the density model to gradients of the wave function.  
Score matching is a best-in-class algorithm for learning \emph{local} features of the density, but it is known to struggle to model a density with multiple modes that are separated by large regions of low probability \citep{song2019generative}, which we encounter in the case of molecular dissociation.
To mitigate this, the second loss term uses noise-contrastive estimation \citep{gutmann2010noise}, which yields well-calibrated \emph{global} features of the learned electron density, such as charges on dissociated molecular fragments.

\methodabb\ is only as accurate as the underlying wave functions and in this work we chose to couple it to wave functions from VMC with deep-learning ansätze (deep QMC).
Deep QMC is a recent class of ab-initio electronic structure methods that model the many-electron wave function $\Psi$ using a suitable neural network \citep{han_solving_2019,hermann2020deep,pfau2020ab}, see\phantom{-}\citet{hermann2023ab} for an in-depth review. Training this network relies only on the variational principle and requires no data as an input, as the electron configurations are generated by Monte Carlo sampling during training. This approach does not use a Gaussian basis, but instead learns a much more flexible neural network model of the wave function, and as such offers one of the most accurate many-electron wave-function models in existence.

\begin{figure*}[tpb]
    \centering
    \includegraphics[width=1.8\columnwidth]{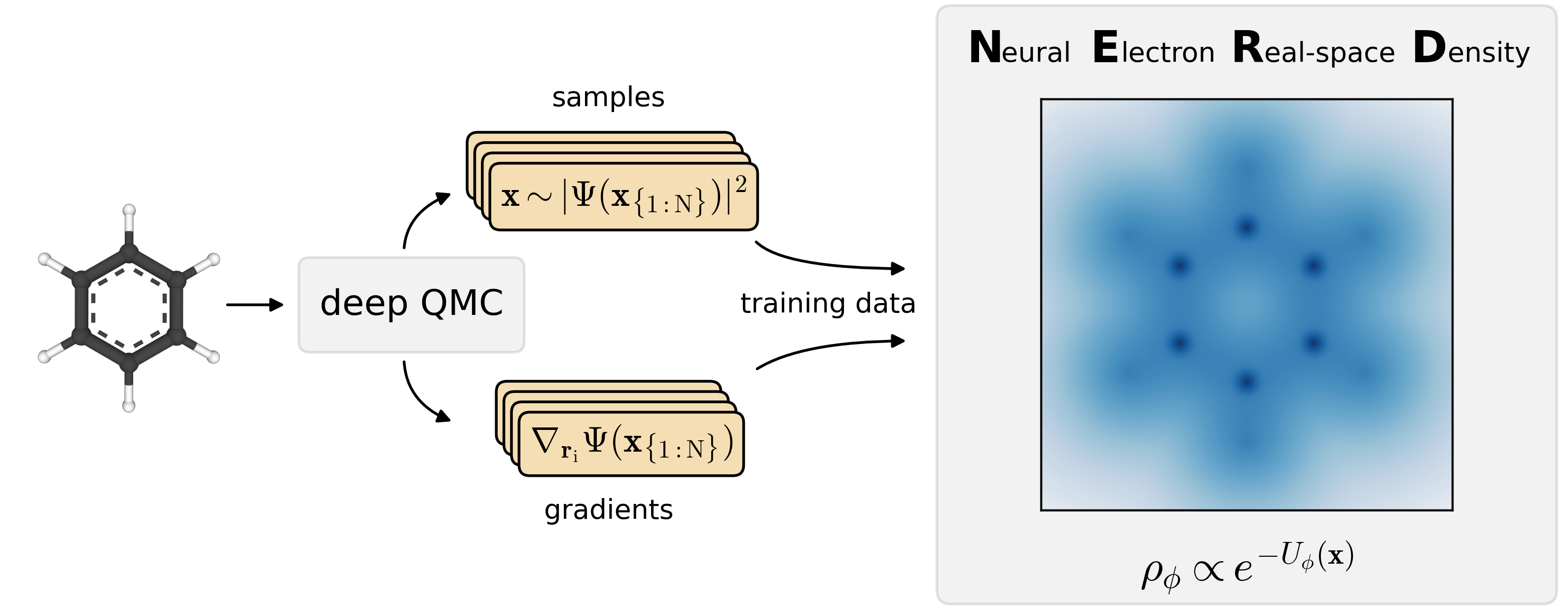}
    \caption{Schematic diagram for training of {\methodabb}s. For an input molecular geometry a highly accurate wave function is obtained with deep QMC. Samples from the wave function and associated gradients serve as training data in a combined score matching and noise contrastive estimation approach to optimizing the {\methodabb}.}
    \label{fig:diagram}
\end{figure*}

To analyze and verify the densities from \methodabb\ models, we first compare them on small molecules to densities from other methods such as full CI (FCI) in a large basis, showing that we capture the same general features but with improved asymptotics.
We then show that for quantities involving \emph{derivatives} of the density, our model is superior to Gaussian-based alternatives.
We also demonstrate that a \methodabb\ model, once trained, delivers very accurate values for one-electron observables through numerical integration. Hence, with one single density model we have access not only to very accurate densities and density derivatives, but also to forces on the nuclei (via the Hellmann--Feynman theorem), spin-densities on the nuclei, dipole moments, etc. This is a crucial difference with respect to traditional QMC evaluation of these observables, which requires different dedicated estimators for each one of them.\cite{assaraf2003zero,hakansson_efficient_2008,per_zero-variance_2011,assaraf_improved_2007,VarBarGui-JCP-2014}
Finally, we investigate the scalability and convergence rate of our density model using benzene as an example.

In summary, we present a method, dubbed \methodabb, to extract one-electron density from real-space many-electron wave functions at high fidelity, and show that it is possible to obtain highly accurate one-electron densities from deep-learning wave function ans\"{a}tze. Figure \ref{fig:diagram} displays the schematic diagram of training {\methodabb}s.

\section{Preliminaries}

\subsection{Variational and deep quantum Monte Carlo}
\label{sec:nnwf}
VMC is a method for approximately solving the Schr\"{o}dinger equation by minimizing the energy expectation value of a parametrized ansatz. We focus on solving the time-independent, electronic Schr\"{o}dinger equation of molecular systems in the Born--Oppenheimer approximation, defined by the position and charge of nuclei $I=1,\dotsc,M$, denoted $\Rv_I, Z_I$.

A first-quantized, real-valued wave function ansatz parametrized by $\theta$ is a function $\Psi_\theta(\xv_1, \xv_2, \dotsc, \xv_N) \in \mathbb{R}$, with $\xv_i = (\rv_i, \sigma_i)$ denoting the spatial-spin coordinates of electron $i$, that is constrained to be antisymmetric under the exchange of any two electrons.
The parameters $\theta$ are trained variationally by minimizing the Rayleigh quotient
\begin{equation}
\label{eq:qmc}
	\theta' = \argmin_{\theta} \frac{\langle \Psi_{\theta} | \hat{H} | \Psi_{\theta} \rangle}{\langle \Psi_{\theta} | \Psi_{\theta} \rangle},
\end{equation}
and the electron configurations $\mathbf{r}_i$ at which (\ref{eq:qmc}) and its derivatives are computed are generated via Monte Carlo sampling from the many-body density $|\Psi_{\theta}|^2$.
After optimization, the ground state energy can be obtained as $E = \langle \Psi_{\theta'} | \hat{H} | \Psi_{\theta'} \rangle / \langle \Psi_{\theta'} | \Psi_{\theta'} \rangle$. 

Traditional VMC uses ansätze built from Slater determinants of precomputed single-particle orbitals, and optimizes at most a few thousand wave-function parameters, typically in the form of a symmetric Jastrow factor. Deep QMC introduces wave functions parametrized by neural networks which may depend on millions of parameters and are trained with deep-learning techniques such as stochastic gradient descent. Neural network wave functions were first introduced by Carleo and Troyer\citep{carleotroyer2017}.
In 2020, PauliNet and FermiNet were introduced as the first high-accuracy solutions for molecular ground states, using a first-quantized ab-initio neural network approach \citep{hermann2020deep,pfau2020ab}. Various new network architectures and algorithmic improvements have since been made to further improve accuracy and efficiency \citep{spencer_better_2020,SchatzleJCP23,gao2024neural,li2024computational,gerard_2022,lin_explicitly_2023,kim2024neural, li_fermionic_2022,glehn2022self}.

Most neural network wave functions for molecules employ
generalized Slater determinants, that augment the single-particle orbitals of conventional Slater determinants with many-body correlation,

\begin{align}\label{eq:dl_ansatz}
  \Psi_{\boldsymbol\theta}(\xv_{1:N}) &= \textstyle{\sum}_p c_p
  \det[\mathbf A_\uparrow^p(\xv_{1:N})]\det[\mathbf A_\downarrow^p(\xv_{1:N})] \, , \\
    {A_\sigma^p}_{ik}(\xv_{1:N})&= \phi^p_{\sigma k}(\rv^\sigma_i, \{\rv^\uparrow\},\{\rv^\downarrow\}) \times \varphi^p_{\sigma k}(\rv^\sigma_i) \, , \label{eq:mb_orbital}
\end{align}
where $\xv_{1:N}=\{\xv_1, \xv_2, \dotsc, \xv_N \}$ and $\mathbf r_{1:N}=\{\mathbf r_1, \mathbf r_2, \dotsc, \mathbf r_N \}$ denote the combination of all the spatial-spin coordinates and electron configurations, respectively.
Here, electrons are grouped into (permutation invariant) sets of spin-up and spin-down electrons, $\{\mathbf r^\uparrow\}$, $\{\mathbf r^\downarrow\}$, $\sigma =~\uparrow$ or $\downarrow$ to denote different groups of spins, $\phi^p_{\sigma k}$ are many-body spin orbitals, and $\varphi^p_{\sigma k}$ are single-particle spin envelopes that ensure the correct asymptotic behavior of the wave function with increasing distance from the nuclei. 
The ansatz may be a linear combination of multiple generalized Slater determinants, which are distinguished with the $p$ index. The form of $\phi^p_{\sigma k}$ in \eqref{eq:mb_orbital} is closely related to the backflow transformation~\cite{feynman_energy_1956,CeperleyJSP91}, which introduces quasi-particles to obtain many-body orbital functions.
The key observation motivating this choice of many-body orbitals is that the antisymmetry of Slater determinants constructed from many-body orbitals is preserved as long as the orbital functions are equivariant with the exchange of electrons.

For further details on VMC with neural network wave functions, see\phantom{-}\citet{hermann2023ab}
In all experiments of the present paper, the Psiformer wave-function architecture is used, the details of which are described in\phantom{-}\citet{glehn2022self} .

\subsection{Electron density}
\label{sec:electron_density}

In practical VMC one works with the unnormalized wave function $\Psi$, from which
the electron spin-density is derived as the one-body marginal of $\abs{\Psi}^2$ normalized to the number of electrons $N$,
\begin{align}
\label{eq:density_definition}
\rho(\xv) = \frac{N}{\langle \Psi | \Psi \rangle} \int \dd \xv_{2\dotsc N} \abs{\Psi(\xv_1, \xv_2, \dotsc, \xv_N)}^2.
\end{align}

There are two known key properties of the electron density.
\paragraph{Kato's cusp condition}
Kato's cusp condition \citep{kato1957eigenfunctions} states that for each nucleus, the electron density has a non-differentiable `cusp' at the point $\Rv_I$ such that\footnote{We express these conditions in terms of $\nabla \log \rho$ as it gives the cleanest comparison with the density models that we introduce in Section~\ref{sec:method}.}
\begin{equation}
\label{eq:cusp_condition}
\frac{1}{4\pi}\lim_{\epsilon \downarrow 0} \int_{\rv \in S^2} \rv \cdot \nabla \log \rho(\Rv_I + \epsilon \rv, \sigma) = -2Z_I.
\end{equation}
\paragraph{Exponential decay and tail convergence}
Away from the nuclei, $\rho$ should be differentiable \citep{Fournais2002densityanalytic, Fournais2004densityanalytic} and the tail of the distribution decays exponentially. Specifically, 
\begin{equation}
\label{eq:ip}
\log \rho(\rv,\sigma) \leq -2\kappa_{\sigma}\|\rv\| + 2\beta_{\sigma}\log(1+ \|\rv\|) + \text{const.},
\end{equation}
where \(\kappa_\sigma = \sqrt{2\mathrm{IP}(\sigma)}\) with 
$\mathrm{IP}(\sigma) > 0$ as the ionization potential and \(\beta_{\sigma} = \left(\sum_I Z_I-N+1\right)/\kappa_\sigma - 1\) \citep{HoffmanOstenhof1977densityDecay, Ahlrichs1981densityDecay}. 
Unlike with Kato's cusp condition, $\mathrm{IP}(\sigma)$ requires calculation to obtain, so we treat this constraint as only guiding the functional form of the model.

\subsection{Previous developments on classical VMC density estimators}
Significant progress has already been made in developing low-variance and/or low-bias estimators for evaluating the VMC electron density.
The most commonly used density estimator is the Assaraf--Caffarel--Scemama zero-bias estimator (ACS-ZB) \cite{assaraf_improved_2007,VarBarGui-JCP-2014}, which is general and unbiased. However, the ACS-ZB estimator has a formally divergent variance due to a quadratic divergence caused by the square of the drift velocity term. This results in a non-integrable quartic divergence when an electron approaches the nodal surface, leading to an inaccurate density at the nucleus.\cite{VarBarGui-JCP-2014} To reduce variances, careful designs of different auxiliary functions in ACS-ZB for the points located close to a nucleus or at tails are also proposed.\cite{VarBarGui-JCP-2014} The equation of ACS-ZB estimator and the proposed auxiliary function designs are shown in Supporting Information Sec. S6.
To fully resolve this divergent variance issue of ACS-ZB at nucleus, Ref. ~\citenum{per_zero-variance_2011} proposes a Per--Snook--Russo zero-variance-approximate-zero-bias estimator (PSR-ZV(ZB)) specifically for the density evaluation at nucleus. Although the PSR-ZV(ZB) estimator provides high quality density estimates at the nucleus, it lacks generality and cannot be applied at other positions. The details of PSR-ZV(ZB) estimator are listed in Supporting Information Sec. S6.
Compared to \methodabb, the most significant drawback of all these traditional density estimators is that they only provide an estimate of the density at a single point, and therefore must rely on a grid to represent the complete, three dimensional density \citep{assaraf_improved_2007,VarBarGui-JCP-2014,hakansson_efficient_2008,per_zero-variance_2011}.
Not only does this increase their practical computational costs vastly, it also makes it difficult to reuse previously computed densities for new tasks, which might require a different level or type of grid.

\subsection{Machine-learning energy based models}
\label{sec:ebms}
In machine learning, energy\footnote{The term `energy' in EBMs is not connected to the energy in quantum chemistry that arises as an eigenvalue of the Hamiltonian. The `energy' in our setting is equivalent to $-\log \rho$ and thus we use a notation based on $U_\phi \approx -\log \rho + C$} based models (EBMs) are a highly flexible class of probability density models that focus on learning an approximation to the logarithm of the unnormalized density of a target distribution. An EBM with parameters $\phi$ over a variable $\xv$ has a probability density
\begin{equation}
    \label{eq:ebm}
    p_\phi(\xv) = \frac{e^{-U_\phi(\xv)}}{\int e^{-U_\phi(\xv')} d\xv'},
\end{equation}
where $U_\phi$ is a real-valued function (e.g.~a neural network).
Due to the intractability of the normalizing constant in Eq.~\eqref{eq:ebm}, na\"{i}ve maximum likelihood training of EBMs is not possible. Numerous training approaches have been proposed\citep{song2021train}, including contrastive divergence \citep{hinton2002training}, score matching \citep{hyvarinen2005estimation,vincent2011connection}, and noise-contrastive estimation \citep{gutmann2010noise}.

\section{Electron Density Estimation}
\label{sec:method}
\subsection{Designing an EBM for the electron density}
\label{sec:model_design}
We propose an EBM approach to the problem of electron density estimation.
EBMs are a promising model class for this problem because they do not place any direct constraints on the form of the density model, meaning that we can incorporate scientific priors as we see fit to match known asymptotic properties of the electron density.

Specifically, we propose an additive form for the unnormalised log-density
\begin{equation}
    \label{eq:additive_model}
    U_\phi(\xv) = E_\phi(\xv) + M_\phi(\xv) + C(\rv),
\end{equation}
where $E_\phi$ is an envelope, $M_\phi$ is a simple multi-layer perceptron (MLP), and $C$ controls the cusps around the nuclei.

For the envelopes, $E_\phi$, we first define the smoothed distances $s_{I} = \sqrt{1 + \|\rv - \Rv_I\|^2}$.
To capture the long distance exponential tails and to anchor the density around the nuclei, we use a sum of exponentially decaying envelopes
\begin{equation}
E_\phi(\xv) = -\log \left( \sum_I \pi_I(\sigma) e^{-\zeta_I(\sigma) s_I} \right),
\end{equation}
where $\pi, \zeta$ are learned parameters (we learn separate values for different spin channels). 

Secondly, $M_\phi$ is a neural network. The input features to the network are defined in a manner inspired by the Psiformer, as a concatenation of four-dimensional featurelets, one for each nucleus
\begin{equation}
	\text{input} = \text{concat}_I \begin{pmatrix}
	\log(1 + s_I) \\
	(\rv - \Rv_I) \log (1 + s_I) / s_I
	\end{pmatrix}.
\end{equation}
The network then consists of a four-layer MLP with skip connections, SiLU activations \citep{hendrycks2016gaussian} and with each hidden layer having size 64.
There are two output heads for the up- and down-spin channels.

Finally, the cusp term is
\begin{equation}
\label{eq:cusp_model_form}
C(\rv) = 2\sqrt{\pi}\sum_I  \text{erf}\left(\tfrac{1}{2}Z_I\|\rv-\Rv_I\|\right).
\end{equation}
We can verify that this term will satisfy Kato's cusp condition in Eq.~\eqref{eq:cusp_condition} (see Supporting Information Sec.~S2). Since $E_\phi$ and $M_\phi$ are smooth everywhere, this also means that the model as a whole satisfies Kato's cusp condition.

\subsection{Score matching}
\label{sec:score_matching}

The general principle behind score matching is that, if we can train our model so that $\nabla U_\phi = -\nabla \log \rho$ everywhere, then $e^{-U_\phi} = \rho$ up to a normalizing constant. To achieve this, we could consider minimizing the loss
\begin{equation}
    \label{eq:theoretical_sm}
    \E_{\xv \sim \rho(\xv)} \left[ \| -\nabla_\rv \log \rho (\xv) - \nabla_\rv U_\phi(\xv) \|^2 \right],
\end{equation}
however, this is impractical because it requires us to already know $\nabla \log \rho$.
Instead, we can show that, starting from definition in Eq.~\eqref{eq:density_definition}, we can minimize a loss involving gradients of the \emph{wave function}
\begin{equation}
    \label{eq:wf_sm}
    \E\left[ \frac{1}{N} \sum_{i=1}^N \| -2\nabla_{\rv_i} \log |\Psi(\xv_{1:N})| - \nabla_\rv U_\phi(\xv_i) \|^2 \right],
\end{equation}
where data is sampled from $\xv_{1:N} \sim |\Psi(\xv_{1:N})|^2$ by running Markov Chain Monte Carlo on the deep QMC wave function in the same manner as in QMC training.
A full derivation is presented in Supporting Information Sec.~S1B; the connection to force matching is discussed in Supporting Information Sec.~S1C.
Since we have access to the wave function, we do \emph{not} require a noising--denoising approach and can therefore target the exact electron density.

Score matching is a best-in-class algorithm for learning \emph{local} features of the density, however, it is known to struggle to model a distribution with multiple modes that are separated by a large region of low probability \citep{wenliang2019learning,song2019generative}. Because score matching is based only on the gradient of the log-density, all modes have to be connected by regions of positive density to correctly determine relative weights in multiple modes. However, for a distant system, the target density is nearly zero between two modes, leading to no or few samples in between. Therefore, score matching has insufficient evidence (only a very weak signal) to correctly calibrate the relative masses of several far-separated modes.

\subsection{Noise contrastive estimation}
\label{sec:nce}
To correct the relative masses of separated modes, we seek a loss function that directly trains the values of the log-density. Noise contrastive estimation \citep[NCE]{gutmann2010noise} is one such approach.
In NCE, we set up a synthetic classification problem to distinguish samples from the true distribution, $\rho$, and from a noise distribution, $p_n$, which has a known density.
In our case, we sample the joint distribution $y, \xv \sim p(y=1)\rho(\xv)+p(y=0)p_n(\xv)$ where $p(y=1)$ is a free parameter that determines the fraction of true and noise samples.
Samples from $\rho$ are obtained by sampling $|\Psi|^2$.
We then set up a `classifier' model to predict $y$ given $\xv$ with probability
\begin{equation}
	p_\phi(y=1 \mid \xv) = \frac{e^{-U_\phi(\xv)}}{e^{-U_\phi(\xv)} + \nu p_n(\xv)},
\end{equation}
where $\nu = p(y=1)/p(y=0)$,
and then minimize the negative log-likelihood loss on our observed data
\begin{equation}
\label{eq:nce_loss}
    \E_{y, \xv \sim p(y=1)\rho(\xv)+p(y=0)p_n(\xv)}[-\log p_\phi(y \mid \xv)].
\end{equation}
When the model is sufficiently powerful, the trained classifier will match the Bayes-optimal classifier, with
\begin{equation}
	p(y=1\mid \xv) = \frac{\rho(\xv)}{\rho(\xv) + \nu p_n(\xv)},
\end{equation}
requiring $e^{-U_\phi} = \rho$. 

The success of NCE is known to depend heavily on the choice of noise distribution\citep{chehab2022optimal}. For example, $p_n$ should be at least as heavy-tailed as the data.
In the quantum chemistry context, we use the known exponential tails of the density, and the fact that electron density should cluster near nuclei, to define a suitable noise distribution.
Our noise distribution, $p_n(\xv)$, is a weighted mixture of exponential-tailed distributions centered on atoms,
\begin{equation}
p_n(\xv) \propto{ \sum_I Z_I e^{-\|\rv - \Rv_I\|}}.
\end{equation}
Combining the score matching loss (Eq.~\eqref{eq:wf_sm}) with the NCE loss (Eq.~\eqref{eq:nce_loss}) gives a training mechanism to accurately capture both local and global features of the target density. Our overall loss function used during training is therefore
\begin{equation}
    \ell = \ell_\textrm{SM} + \lambda \ell_\textrm{NCE},
\end{equation}
where we use $\lambda=1$ unless otherwise stated.

\section{Results}
Detailed experimental settings for the QMC calculations and the \methodabb\ trainings are listed in Table S1 and Table S2 (Supporting Information Sec. S9), respectively. 

\subsection{Small atoms}

As the first set of experiments, we assess the quality of the deep QMC wave functions and subsequently that of the \methodabb\ models, on a set of small atoms ranging from \ch{He} to \ch{Ne}.
First, the accuracy of the many-body wave functions is validated by comparing deep QMC ionization potentials (IPs) with reference experimental values \cite{nistNISTAtomic}.
The QMC ionization potential values are computed as $E_\text{IP} = E_\text{cation}-E_\text{atom}$, where $E_\text{atom}$ and $E_\text{cation}$ are the energy expectation values of the trained Psiformer ansätze for the atom and cation, respectively. 
The computed Psiformer ionization potentials agree with experimental results extremely well, with percentage errors ranging from 0.029\% to 0.37\%, indicating the high accuracy of the corresponding wave functions.

\begin{table}[tbp]  
\centering  
\caption{Ionization potentials (IPs) of first row atoms from experimental results\citep{nistNISTAtomic} and deep QMC calculations. All energy values are in the unit of eV.}
\label{tab:atoms_ip}
\begin{ruledtabular}
\begin{tabular}{lddd}  
Atom &\multicolumn{1}{c}{Exp. IP} & \multicolumn{1}{c}{deep QMC IP} & \multicolumn{1}{c}{Error\%}  \\
\hline  
He & 24.587 & 24.594  & 0.029\%\\
Li & 5.3917 & 5.3873  & -0.082\%\\
Be & 9.3227 & 9.3576  & 0.374\%\\
B  & 8.2980 & 8.3047  & 0.081\%\\
C  & 11.260 & 11.238  & -0.195\%\\
N  & 14.534 & 14.563  & 0.200\%\\
O  & 13.618 & 13.656  & 0.279\%\\
F  & 17.423 & 17.438  & 0.086\%\\
Ne & 21.565 & 21.608  & 0.199\%\\
\end{tabular}
\end{ruledtabular}
\end{table}

\begin{table}[tbp]  
\centering  
\caption{Contact densities (densities at nuclei) from Slater-Jastrow VMC with PSR-ZV(ZB) estimator\cite{per_zero-variance_2011}, CCSD/aug-cc-pVQZ wave functions, and our \methodabb\ model. All the values are in atomic units (a.u.).} 
\label{tab:atoms_density}
\begin{tabular}{l|ccc}
\hline  
\hline  
Atom & \multicolumn{1}{c}{VMC PSR-ZV(ZB)} \citep{per_zero-variance_2011} & \multicolumn{1}{c}{CCSD/aug-cc-pVQZ} &  \multicolumn{1}{c}{\methodabb} \\
\hline  
He & ---    & 3.3830 & 3.6503 \\
Li & 13.847 & 13.467 & 13.914 \\
Be & 35.540 & 34.616 & 35.625 \\
B  & 72.242 & 70.423 & 72.517 \\
C  & 128.16 & 124.96 & 128.76 \\
N  & 207.02 & 201.99 & 207.66 \\
O  & 312.75 & 305.80 & 314.24 \\
F  & 450.64 & 439.79 & 452.64 \\
Ne & 622.50 & 608.61 & 624.49 \\
\hline  
\hline  
\end{tabular}  
\end{table} 

\methodabb\ models are then trained for each atom using the verified Psiformer wave function. As a first quality indicator of  \methodabb, we show that it yields accurate estimates for the electron densities at the position of the nuclei, also known as \emph{contact density}. Contact density is linked to various experimental techniques for probing the local chemical environment of a nucleus, such as M\"o\ss bauer isomer shift in M\"o\ss bauer spectroscopy \cite{filatov2012analytic} and the Fermi-contact contribution to the hyperfine coupling in electron paramagnetic resonance spectroscopy  \cite{hedegard2013validating}.
In spite of its significance, obtaining accurate contact densities continues to be a challenging task in electronic structure theory \citep{hakansson_efficient_2008,per_zero-variance_2011}.
Densities computed using theories involving Gaussian basis functions have notoriously large errors due to the incorrect cusp behavior of Gaussians.
Traditionally, accurate contact densities are also difficult to extract from real-space QMC wave functions, prompting the development of several dedicated estimators \citep{hakansson_efficient_2008,per_zero-variance_2011}. Here we highlight that {\methodabb} yields accurate densities everywhere in one shot, including at the nuclei. Although traditional ACS-ZB estimator can provide a general way to evaluate densities, it requires independent evaluation for each point and suffers the divergent variance issue when an electron approaches a nodal surface. The dedicated PSR-ZV(ZB) estimator is accurate but can only evaluate the density at the nucleus (see Supporting Information Eq. S48).
In fact, as shown in Table \ref{tab:atoms_density}, the obtained \methodabb\ values are close to the contact densities obtained in Ref.~\citenum{per_zero-variance_2011} from a Slater-Jastrow VMC wave function with PSR-ZV(ZB) estimator.
For comparison, we also report CCSD contact densities computed in-house with the relatively large basis set of aug-cc-pVQZ, which, as expected from the lack of nuclear cusp, consistently yields much lower values.

\begin{figure}[tbp]
    \centering
    \includegraphics[width=0.95\columnwidth]{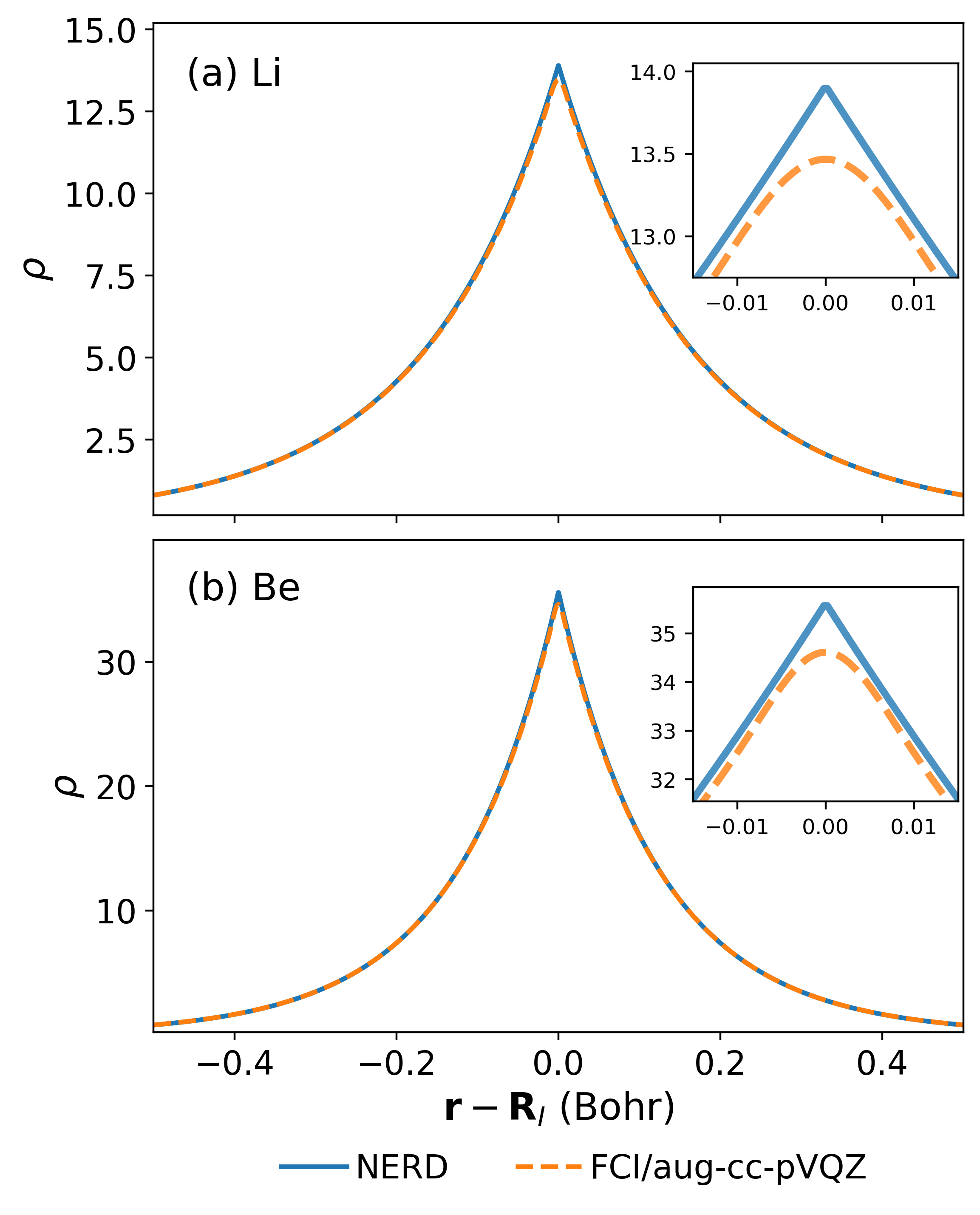}
    \caption{Comparison between densities from the neural network model and FCI/aug-cc-pVQZ for (a) Li atom and (b) Be atom. Both nuclei are located at the origin. Inset plots are displayed to better visualize the cusps around nuclei.}
    \label{fig:li_be_small_r}
\end{figure}

\begin{figure}[tbp]
    \centering
    \includegraphics[width=0.95\columnwidth]{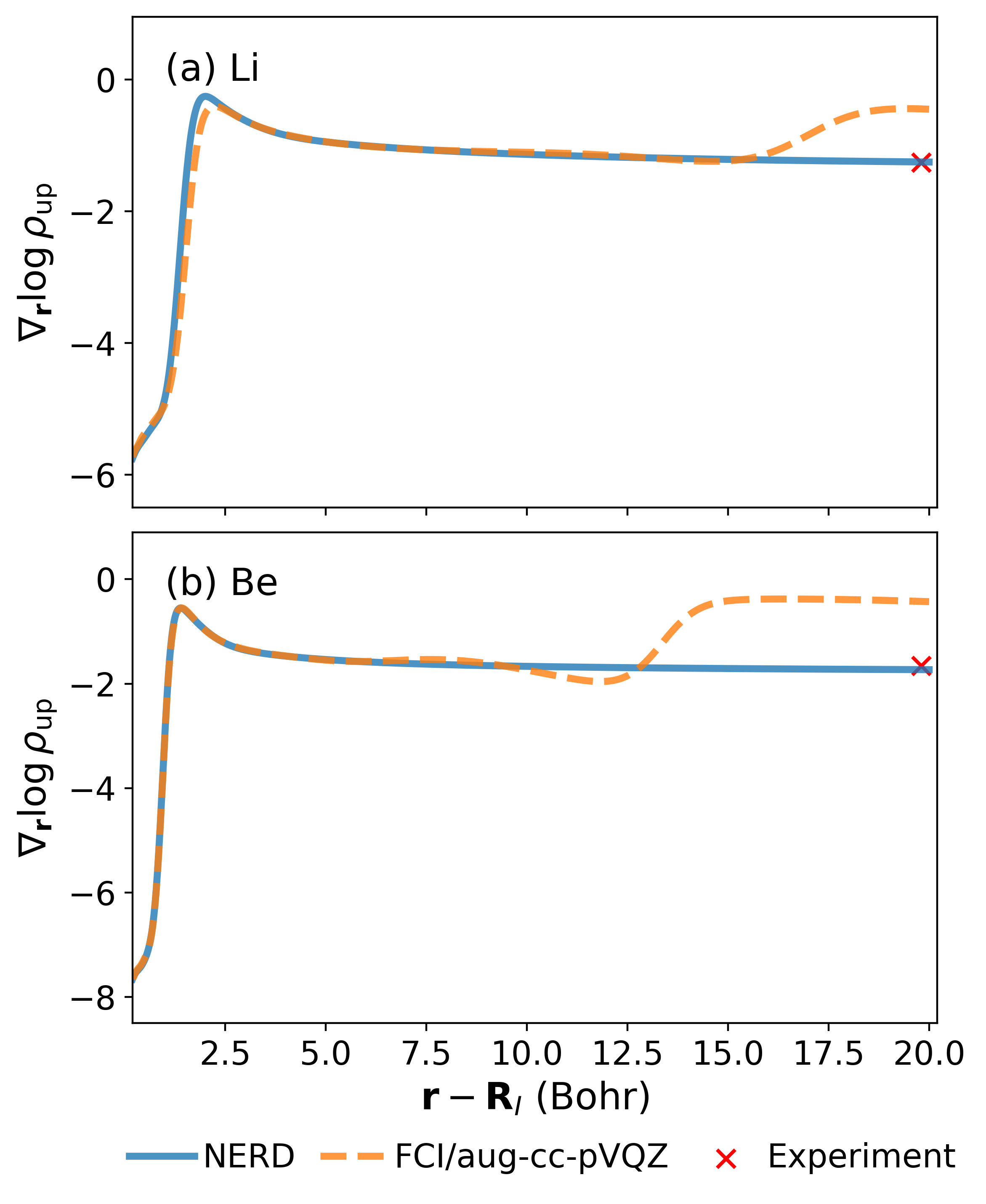}
    \caption{Comparison between radial derivatives of log of spin-up densities from the neural network model, FCI/aug-cc-pVQZ and converting literature ionization potential values via Eq.~\eqref{eq:ip} for (a) Li atom and (b) Be atom. Both nuclei are located at the origin.
    }
    \label{fig:li_be_large_r}
\end{figure}

Next, three key features of the obtained densities are observed at different radial distance scales. Without specification, all the displayed density are  plotted along the $x$-axis in real-space (evaluated by fixing the $y$ and $z$ coordinates).
First, the behaviour of \methodabb\ and FCI/aug-cc-pVQZ densities are compared close to the nucleus.
Following Eq.~\eqref{eq:cusp_condition}, it is known that electronic densities should approach exponentials around the nuclear cusp, with a non-differentiable point at the nucleus.
As is clear from the inset plots of Fig. \ref{fig:li_be_small_r}, densities from \methodabb\ models for the Li and Be atoms have exactly this behaviour, while FCI/aug-cc-pVQZ densities exhibit smooth local maxima instead due to the atom-centered Gaussian basis functions.
In NERD, the carefully crafted nuclear cusp term of Eq.~\eqref{eq:cusp_model_form} ensures the correct behaviour of the {\methodabb}s. In Supporting Information Sec. S7B, we additionally verify the correct cusp using effective potential in Fig. S1 and discuss the accuracy of spin-up and spin-down {\methodabb}s for \ch{Li} and \ch{Be} in Fig. S2. 

Second, the correct exponential decay of the  \methodabb\ model is highlighted in Fig.~\ref{fig:li_be_large_r}, showing  $\nabla_\rv \log \rho(r)$ as a function of $r$ for the Li and Be atoms.
According to Eq.~\eqref{eq:ip},  $\nabla_\rv \log \rho(r)$ should approach the constant $-2 \kappa_\sigma =-2\sqrt{2I_{\mathrm{P}}(\sigma)}$ when $r\to\infty$, marked with red crosses on Fig. \ref{fig:li_be_large_r}.
Comparing  {\methodabb}s with those obtained from FCI/aug-cc-pVQZ calculations, it is clear that the former converges to the correct asymptotic value, while the latter does not.
The incorrect behaviour of the FCI/aug-cc-pVQZ density is rooted in its use of Gaussian basis functions, with which the correct exponential decay cannot be expressed.
In contrast, our  \methodabb\ models are constructed using exponential envelopes, which enables them to correctly describe densities far away from nuclei.

\begin{figure}[tbp]
    \centering
    \includegraphics[width=0.95\columnwidth]{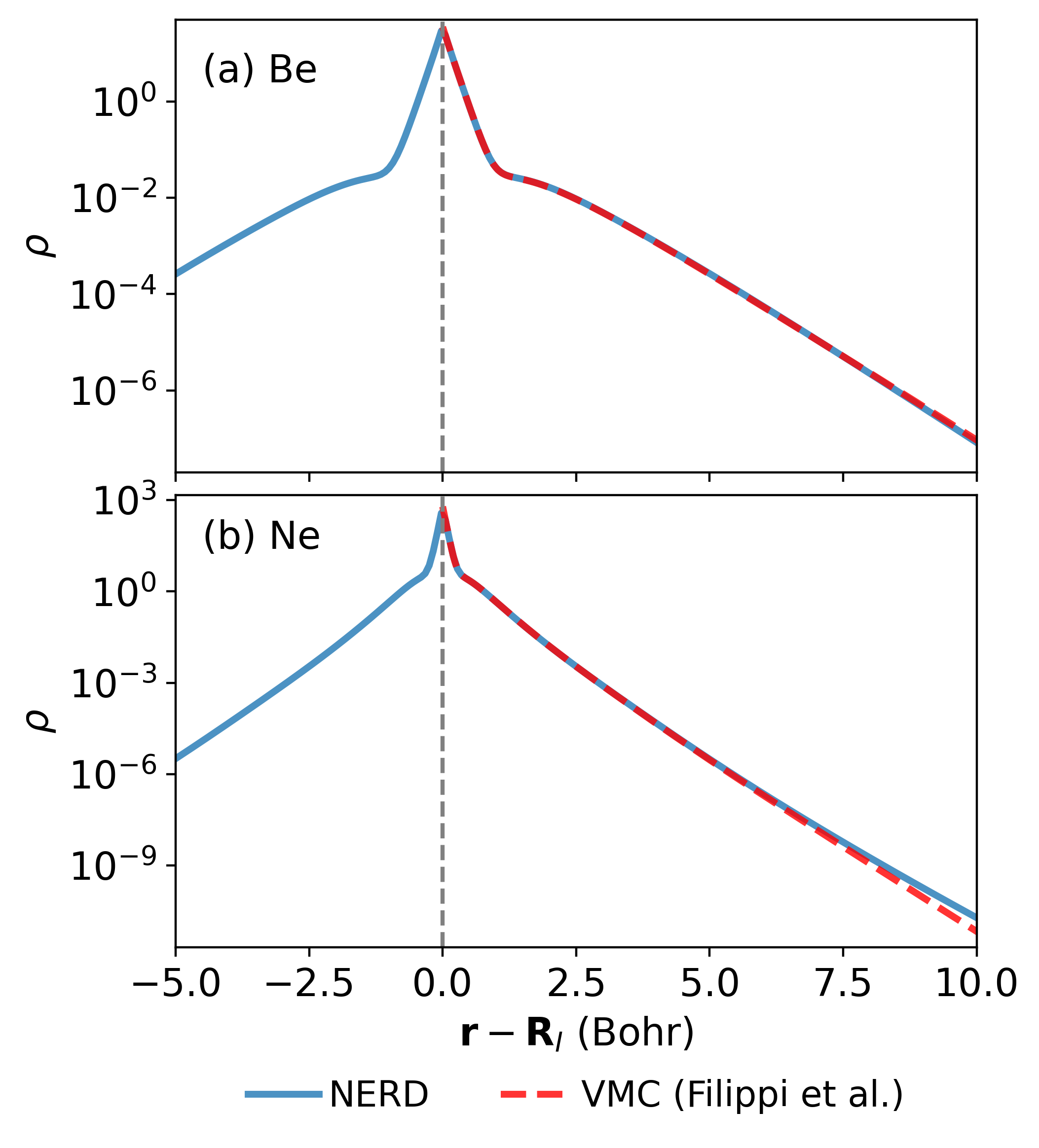}
    \caption{Comparison between the total densities from the neural network model and literature for (a) Be atom and (b) Ne atom. Both nuclei are located at the origin and indicated by the gray dashed lines. The y-axis is plotted on a log scale. The literature densities are computed using VMC and obtained from Filippi et al. \citep{filippi1996generalized}}
    \label{fig:lit_rho_be_ne}
\end{figure}

Third, the reliability of the \methodabb\ models is investigated at intermediate distances from the nucleus.
Figure \ref{fig:lit_rho_be_ne} compares our  \methodabb\ models with highly accurate references obtained with specialized FCI-based methods \citep{filippi1996generalized} for the Be and Ne atoms.
At this intermediate range where we expect both methods to work well, QMC densities agree well with literature values, validating their soundness.

\begin{figure*}[tbp]
    \centering
    \includegraphics[width=1.65\columnwidth]{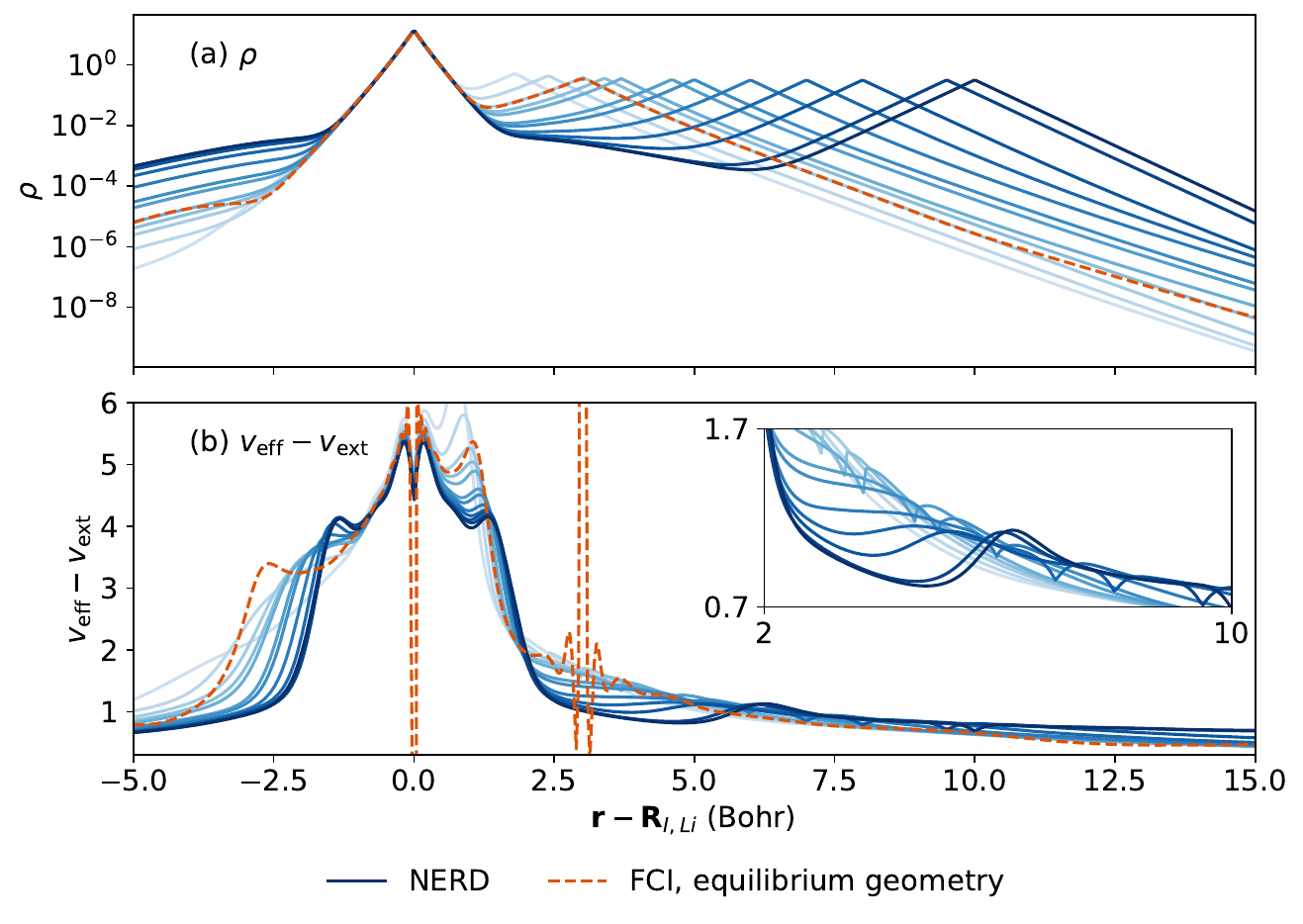}
    \caption{(a) NERD and FCI/aug-cc-pVTZ densities along the dissociation of the \ch{LiH} molecule, (b) effective potentials (see eq.~\eqref{eq:veff}) extracted from the densities. The densities in (a) are plotted on a logarithm scale for y-axis. Darker shades of blue correspond to longer bond lengths, while the FCI/aug-cc-pVTZ calculation was carried out at equilibrium bond length only.}
    \label{fig:lih_ks}
\end{figure*}

\subsection{\ch{LiH} dissociation} \label{sec:lih}
Next, the dissociation of the \ch{LiH} molecule is investigated, by extracting QMC densities for a total of twelve molecular geometries, with bond lengths varying between 1.8 and 10 Bohr.
Cuts from these densities along the bond axis are shown on the top panel of Fig. \ref{fig:lih_ks}, using darker shades of blue for longer bond lengths, with the \ch{Li} atom centered at the origin, and the \ch{H} atom shifting gradually to the right.
From this plot, one can immediately verify that the magnitudes of the density peaks around the two atoms remain constant across all bond lengths.
This behavior is the result of including the NCE loss term during density fitting, ensuring the correct estimation of the relative magnitudes of separated density modes.
The FCI/aug-cc-pVTZ density of \ch{LiH} at its equilibrium geometry is shown on the top panel of Fig.~\ref{fig:lih_ks} with the orange dashed line.
It is in excellent agreement with the density extracted from QMC in the high density region around the nuclei, and only deviates from it appreciably in areas of low density, farther from the nuclei.
This mismatch between the tails of the two densities can again be attributed to the use of a Gaussian basis set in the FCI calculation, with which the correct, exponentially decaying tails cannot be expressed.

To further assess the quality of the {\methodabb}s, they are used to compute the so called effective potential
\begin{equation} \label{eq:veff}
    v_\mathrm{eff}(\xv) = \frac{1}{8} \|\nabla_\rv \log \rho(\xv)\|^2 + \frac{1}{4} \nabla^2_\rv \log \rho(\xv).
\end{equation}
This potential is an important quantity appearing in the effective Schr\"{o}dinger equation for the square root of the density \citep{hunter1975conditional}.
More importantly for the present work it exhibits some well known, yet traditionally hard-to-capture features during the dissociation of \ch{LiH}, the proper description of which constitutes an exceedingly high benchmark of accuracy for  \methodabb\ models \citep{gritsenko1996effect,baerends1997quantum}.
Among these features is the formation of a step with a slight overshoot between the two atoms as the bond length increases, which aligns the ionization potentials of the two atoms (see, e.g., Ref. \citenum{gritsenko1996molecular, tempel2009revisiting}).

The bottom panel of Fig.~\ref{fig:lih_ks} depicts the effective potential extracted from QMC densities at various bond lengths, with the contribution of the external potential (Coulomb potential of nuclei) removed.
The exact numerical recipe for accurately extracting the $v_\text{eff} - v_\text{ext}$ potential from our  \methodabb\ models is described in Supporting Information Sec.~S3.
Due to the carefully selected functional form of our model, the obtained effective potential precisely cancels the diverging external potential at the nuclear cusps, resulting in the bounded, continuous blue curves on the bottom panel of Fig.~\ref{fig:lih_ks}.
This is in stark contrast to the effective potential derived from the FCI/aug-cc-pVTZ density at the equilibrium bond length, shown in orange.
Due to the Gaussian basis set, there are subtle errors in FCI density close to the nuclei, therefore the derived effective potential cannot exactly cancel the external potential, leading to oscillations and divergences in the dashed orange curve at 0 and 3 Bohr.

The inset plot in the bottom panel of Fig.~\ref{fig:lih_ks} highlights the region between the two atoms and shows the gradual formation of the expected step feature as the bond length increases.
The correct description of this elusive feature is made possible by the precisely reproduced exponential decays of the two density modes, and would therefore be extremely challenging using Gaussian basis sets.

\begin{figure}[tbp]
    \centering
    \includegraphics[width=0.9\columnwidth]{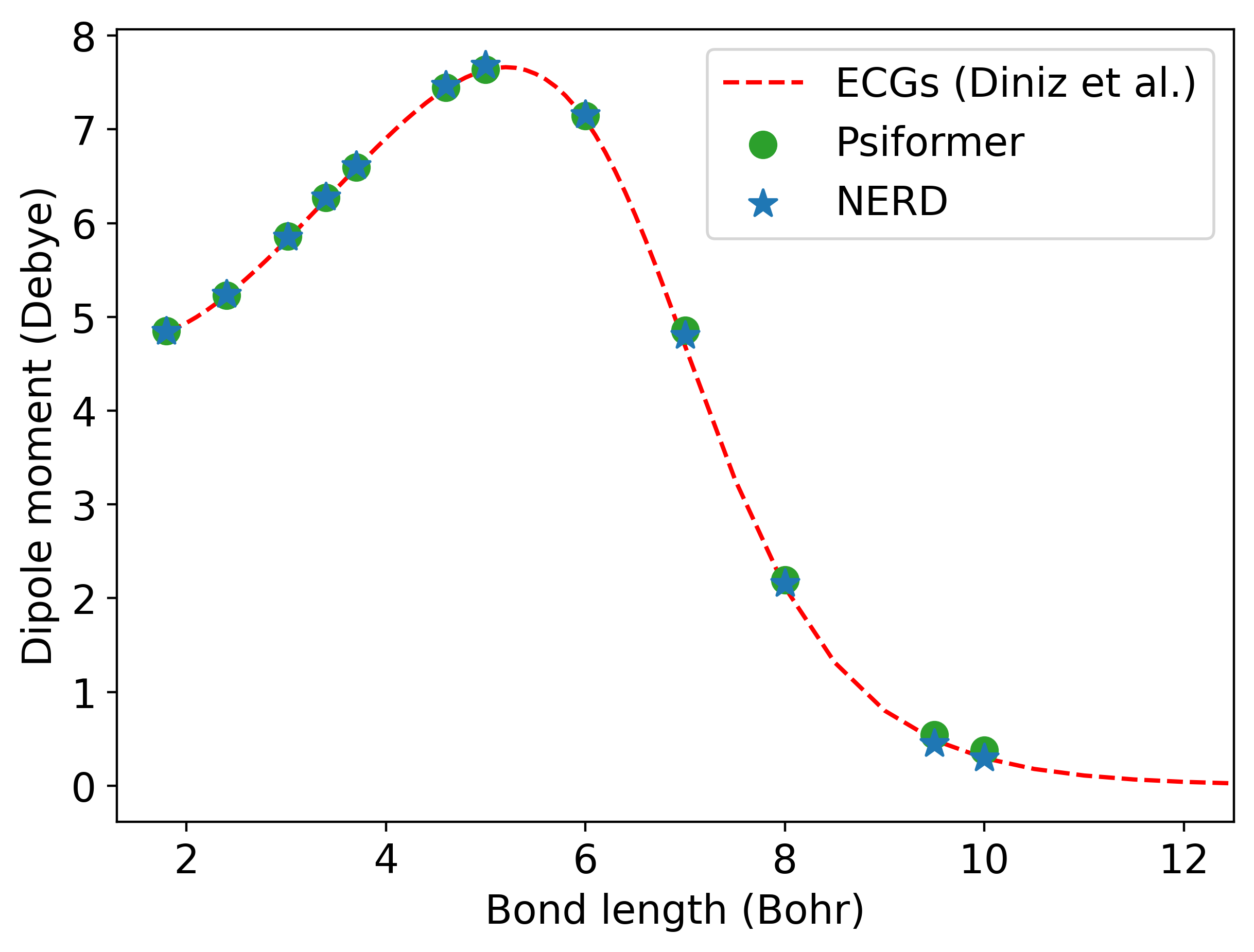}
    \caption{Dipole moments along the dissociation of the LiH molecule. Results evaluated from NERDs are compared with the direct evaluation of the underlying Psiformer wave functions and accurate reference calculations of\phantom{-}\citet{diniz2016accurate} based on explicitly correlated Gaussians (ECGs) with shifted centers. }
    \label{fig:lih_dipole}
\end{figure}

We additionally verify the accuracy of {\methodabb}s by computing the dipole moments, $\bm{\mu}$, from densities and Psiformer wave functions as follows: 
\begin{equation}
    \bm{\mu} = \int \rv\rho(\xv)\,d\xv,\label{eq:dipole}
\end{equation}
Figure \ref{fig:lih_dipole} displays the computed dipole moments for the same \ch{LiH} systems assessed in Fig. \ref{fig:lih_ks}. Evaluations from \methodabb\ models, evaluations from Psiformer wave functions, and the highly-accurate calculations with all-particle explicitly correlated Gaussian functions with shifted centers (ECGs) from Ref.~\citenum{diniz2016accurate}, have excellent agreement along the entire dissociation curve. This indicates the ability of \methodabb\ models to accurately yield other density-based properties.

\subsection{\ch{H4}}
\begin{figure}[thbp]
    \centering\includegraphics[width=1\columnwidth]{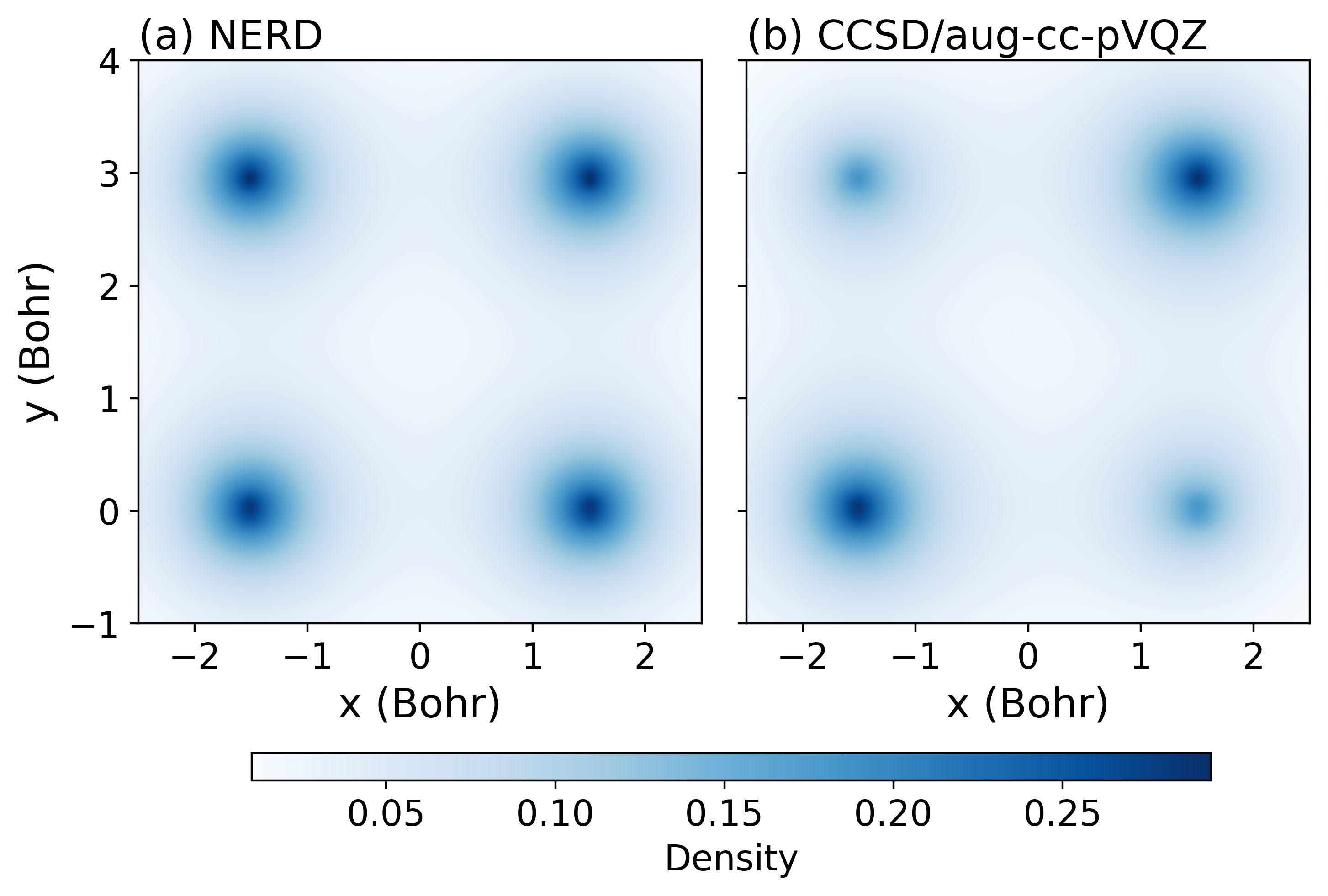}
    \caption{Comparison of spin-up densities computed from (a)  \methodabb\ model using Psiformer wave function and (b) CCSD/aug-cc-pVQZ on a square \ch{H4} molecule with bond lengths of 3.0236 Bohr}
    \label{fig:h4_symmetry_break}
\end{figure}

Deep QMC is known to produce highly accurate results on systems with strong multi-reference character \citep{hermann2023ab}, such as the \ch{H4} molecule in a square configuration  \cite{motta2017towards,rubin2018application}.
To highlight the advantages of QMC densities for these systems, Fig. \ref{fig:h4_symmetry_break} displays planar cuts of \ch{H4} densities obtained from (a) a Psiformer wave function with our  \methodabb\ model, and (b) a CCSD computation in the aug-cc-pVQZ basis.
It is clear that single-reference CCSD yields an incorrect, asymmetric density, as a result of randomly picking one of the two equally important Slater-determinants as its reference determinant.
On the other hand, the density obtained from QMC is perfectly symmetric, verifying the ability of deep QMC to model multi-reference systems and validating the correctness of \methodabb\ model.

\begin{figure*}[tbp]
     \centering
     \includegraphics[width=1\textwidth]{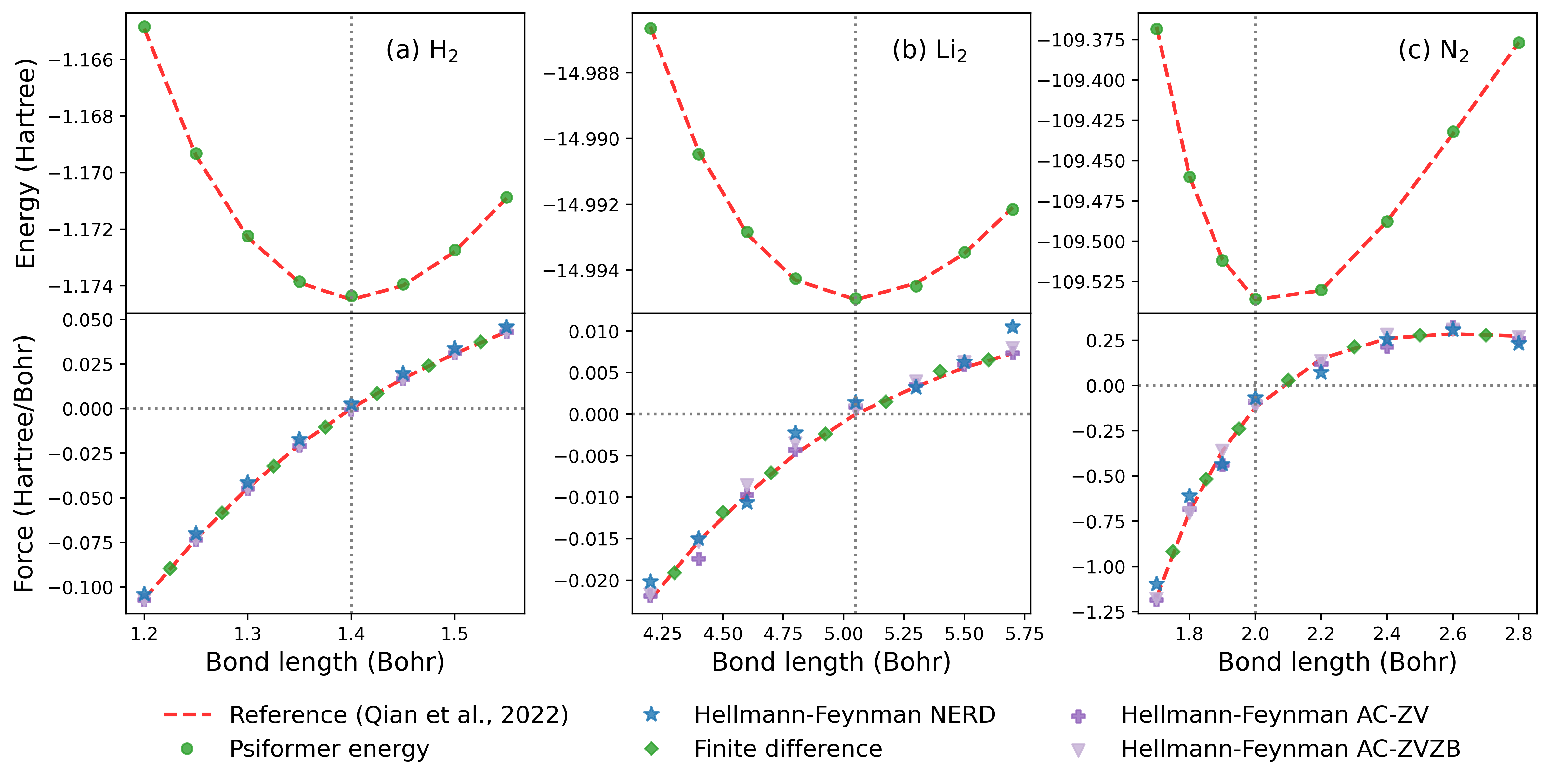}
     
    \caption{Estimating the inter-atomic force for three diatomic systems. The top panels in (a)-(c) show the QMC energies computed from  Psiformer using the settings described in Table S1 in Supporting Information as well as reference energies of\phantom{-}\citet{qian2022interatomic} obtained with FermiNet using 100k training steps for \ch{H2}, and 200 k training steps for \ch{Li2} and \ch{N2}, respectively. In the lower panels, we compare (1) reference forces obtained via FCI \cite{qian2022interatomic} and derivatives of fitted Morse/Long-range  (MLR) potentials \cite{roy2009accurate, roy2006anaccurate}(2) integration of the Hellmann-Feynman force from the density (at 40k density fitting steps), (3) finite difference from QMC energies (4) forces using AC-ZV estimator evaluated on Psiformer (10k evaluation steps), and (5) forces using AC-ZVZB estimator evaluated on Psiformer (10k evaluation steps).
    }
    \label{fig:diatomic_forces}
\end{figure*}

\subsection{Force estimation on diatomic systems}

Here we demonstrate that our highly accurate score matching densities can be used to obtain quantitatively correct Hellmann--Feynman forces \cite{feynman1939forces,politzer2018hellmann}.
To that end, we evaluate energies and Hellmann--Feynman forces for a range of bond lengths of various homonuclear diatomic molecules around their equilibrium geometries.
Specifically, we show that (a) NERD densities are of sufficiently high quality to allow accurate forces to be computed efficiently from them, (b) these forces match the Monte Carlo force estimates from the same wave functions that were used to train the density models, and (c) bias correction schemes accounting for deficiencies in the wave function are not required.

Similar to other observables, the Hellmann--Feynman forces from the  \methodabb\ models are obtained by numerically evaluating the expectation value integrals on a grid (Eq.~S42 in Supporting Information).
While the three-dimensional electron density allows for such treatment, integrals of the $3N$ dimensional wave function have to be estimated through Monte Carlo integration.
Due to the prohibitively high variance of the bare Monte Carlo estimator for nuclear forces, we compare the results with the improved zero-variance Assaraf-Caffarel (AC-ZV) estimator \cite{assaraf2003zero}  (see Supporting Information Sec. S6).
We also test the zero-variance zero-bias Assaraf--Caffarel (AC-ZVZB) estimator (see Supporting Information Sec. S6), which corrects for a potential systematic error due to discrepancies of the wave function from the true ground state, at the cost of additional local energy evaluations.
Furthermore, we compute finite difference forces based on the energy expectation values of our Psiformer wave functions.
This serves as an additional validation of the quality of the wave function in one dimension, but is not a viable strategy on higher dimensional potential energy surfaces.
To validate the forces by comparison with reference results obtained with QMC, we chose the three systems investigated by ~\citet{qian2022interatomic} i.e.~\ch{H2}, \ch{Li2}, and \ch{N2}. 

For all three tested systems, the results presented in Fig.~\ref{fig:diatomic_forces} show that the energies obtained with deep QMC are very close to the reference energies and both the forces extracted from the  \methodabb\ model as well as the Monte Carlo estimates from the wave function are in good agreement with the reference.
Firstly, we find that the integration of the Hellmann--Feynman force on the grid (see Sec. S5 in Supporting Information) gives a good, numerically stable estimate of the force at low computational costs.
Comparing the forces obtained by integrating the density with those obtained directly from the underlying wave function, we find a generally excellent agreement.
With the forces being highly dependent on small changes in density, these results further validate the accuracy of the density fitting procedure.
We attribute the remaining errors to noise in the density fitting as well as a small residual error due to the accuracy of the wave function.
While we conduct the experiment to validate the accuracy of the \methodabb\ models, we find that fitting the density and obtaining forces from the  \methodabb\ model can be considered a workable alternative to the direct force estimation.

The results moreover indicate that the high accuracy of deep QMC wave functions removes the need for bias correction, due to a generally good agreement of the uncorrected forces from both the density and the AC-ZV estimator with the reference.
To further investigate the potential benefits of bias correction we evaluate the AC-ZVZB estimator on the wave functions.
We find that in our experimental setting the bias correction of the AC-ZVZB estimator may not be worth the high cost of additional local energy evaluations, an observation that aligns with the results obtained by ~\citet{qian2022interatomic} .

In conclusion, this experiment shows that our density models provide high quality forces, exemplifying the utility of density extraction and validating the accuracy of the density fitting procedure.
Remarkably, fitting the density on samples from the wave function and evaluating integrals on grids regularizes the high variance of the bare force estimator and gives results in good agreement with the more advanced estimators introduced by Assaraf and Caffarel.
Furthermore, a single  \methodabb\ model suffices to evaluate many observables at a negligible cost compared to the direct estimation of each of them from the wave function by the means of Monte Carlo integration.

\begin{figure*}[tbp]
    \centering
    \includegraphics[width=1.35\columnwidth]{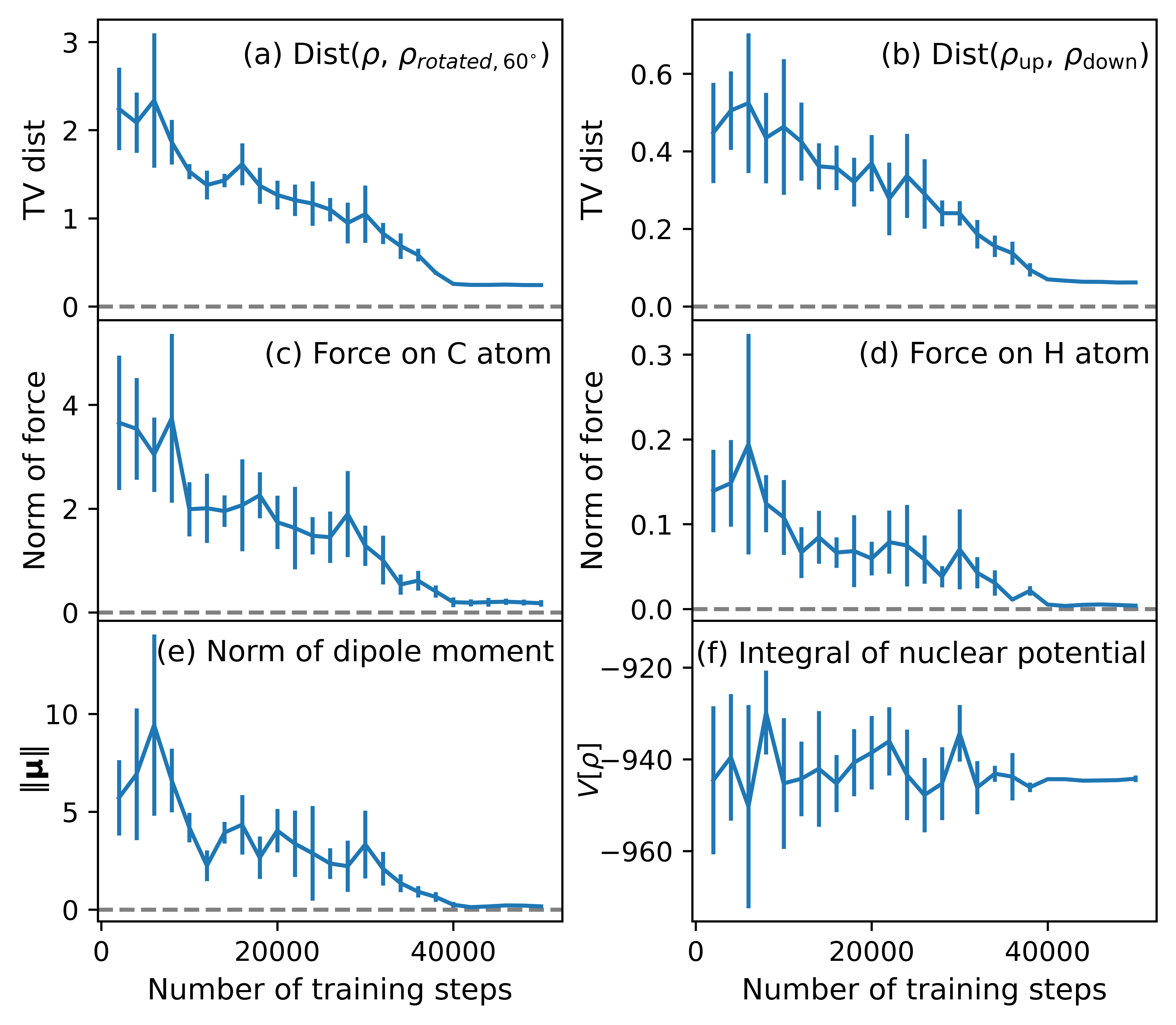}     
     \caption{Evaluating the convergence of our  \methodabb\ model using various metrics on a single Psiformer wave function on the equilibrium benzene. (a) TV distance between $\rho$ and $60^{\circ}$-rotated $\rho$ (b) TV distance between densities of two spins, (c) Hellmann-Feynman force on C, (d) Hellmann-Feynman force on H, (e) Norm of molecular dipole moment, and (f) Integral of nuclear potential. Along the training processes, all the values are computed by averaging results from 5 independent trainings and the error bars are 2 standard deviations of the corresponding means. The gray dashed lines represent the ground truth values. Note that the model learning rate is decayed during training, see Sec. S9 in Supporting Information}
     \label{fig:benzene_stat}
\end{figure*}
\subsection{Benzene}

We finally investigate the scalability and convergence properties of our method on the benzene molecule (with 42 electrons) at its equilibrium geometry. 
To assess the convergence of our model, we trained five  \methodabb\ models with different seeds against the same Psiformer checkpoint. 
At various stages of density training, we computed several metrics and the variance between seeds. To check of the rotation invariance of the density (Fig.~\ref{fig:benzene_stat}a), we compute the total variation (TV) distance \citep{tsybakov2009} between $\rho$ and $60^{\circ}$-rotated $\rho$, which should be zero as the ground state density is invariant under the D\textsubscript{6h} point group. TV distance is a commonly used metric in machine learning to quantify the differences between two probability distributions and its definition is in Supporting Information Eq.~S41. 
We also check on the spin-symmetry of the density (Fig.~\ref{fig:benzene_stat}b) by computing the TV distance between spin-up and spin-down densities, $\mathrm{TV}(\rho_\mathrm{up}, \rho_\mathrm{down})$. This should also be zero for the ground state.
Another metric to check is Hellmann-Feynman force \cite{feynman1939forces,politzer2018hellmann} on one certain atom, which could be computed by Eq.~S42 in Supporting Information. Figure ~\ref{fig:benzene_stat}c and ~\ref{fig:benzene_stat}d display the Hellmann-Feynman force on a carbon and a hydrogen atom, respectively, along the training processes. Both values converge to nearly 0 as expected because the molecule is at its experimental equilibrium geometry.
The dipole moment (Eq.~\ref{eq:dipole}) should also be zero due to the overall symmetry of the system as shown in Fig.~\ref{fig:benzene_stat}e.
Finally, we compute the nuclear potential by Eq.~S43 in Supporting Information  to assess the reduction in variance during training (Fig.~\ref{fig:benzene_stat}f). 
For all the 4 metrics we assessed, as training progresses, the variances between different seeds decrease close to 0 and the values converge close to their theoretical values after $40\,000$ training steps. This investigation suggests that our  \methodabb\ model is accurate and stable with a reasonable convergence speed even for larger systems.

\section{Conclusion}
In this study, we propose an approach, \methodabb, to extract electronic densities from wave functions computed using real-space quantum chemistry methods via score matching and noise contrastive estimation techniques. We first validate the quality of the deep QMC wave functions from Psiformer by showing the agreements between the ionization potential computed from Psiformer energies and experimental values for small atoms. Then, the significant features of the densities, including Kato's cusp condition, tail convergences, and reliability of the densities across the entire space,  are also systematically checked using \ch{Li}, \ch{Be} and \ch{Ne}, as examples.
We also investigate the performances of our {\methodabb}s and dipole moments computed from \methodabb\ models along the dissociation curves of \ch{LiH}. The curvature of FCI densities is incorrect around nuclei while our {\methodabb}s provide the correct solutions. The quality of our \methodabb\ is additionally verified by the resulting physical solutions for two symmetric systems, i.e., \ch{H4}. The \methodabb\ models are also applied to compute 
Hellmann-Feynman forces accurately along the \ch{H2}, \ch{Li2}, and \ch{N2} dissociation curves. Finally, we show the scalability of \methodabb\ by evaluating densities and density-related properties of benzene.
As a general method, \methodabb\ approach paves a way to obtain high-quality densities from any wave function computed by real-space electronic structure theory. As as future direction, we plan to extend applications of the current NERD approach from molecular systems to periodic systems and materials to compute accurate densities and density related properties.

\section*{Data Availability}
The data that support the findings of this study are available from the corresponding author upon reasonable request.

\section*{Acknowledgement}
We thank Professor C. J. Umrigar for providing us the VMC densities of \ch{Be} and \ch{Ne} atoms and Prof. Carlos F. Bunge for providing us the FCI densities with 8s up to k basis functions of \ch{Li}. We also appreciate the useful discussions from Dr. Rianne van den Berg.

\bibliography{biblio}

\end{document}


\title[]{Supporting Information for Highly Accurate Real-space Electron Densities with Neural Networks}
\author{Lixue Cheng}
\thanks{These authors contributed equally to this work.}
\affiliation{Microsoft Research, AI for Science}
\author{P. Bern\'{a}t Szab\'{o}}
\thanks{These authors contributed equally to this work.}
\affiliation{Microsoft Research, AI for Science}
\affiliation{Freie Universit\"{a}t Berlin}
\author{Zeno Sch\"{a}tzle}
\thanks{These authors contributed equally to this work.}
\affiliation{Microsoft Research, AI for Science}
\affiliation{Freie Universit\"{a}t Berlin}
\author{Derk P. Kooi}
\affiliation{Microsoft Research, AI for Science}
\author{Jonas K\"{o}hler}
\affiliation{Microsoft Research, AI for Science}
\author{Klaas Giesbertz}
\affiliation{Microsoft Research, AI for Science}
\author{Frank No\'{e}}
\email{franknoe@microsoft.com}
\affiliation{Microsoft Research, AI for Science}
\affiliation{Freie Universit\"{a}t Berlin}
\author{Jan Hermann}
\email{jan.hermann@microsoft.com}
\affiliation{Microsoft Research, AI for Science}
\author{Paola Gori-Giorgi}
\email{pgorigiorgi@microsoft.com}
\affiliation{Microsoft Research, AI for Science}
\author{Adam Foster}
\email{adam.e.foster@microsoft.com}
\affiliation{Microsoft Research, AI for Science}

\maketitle
\onecolumngrid

\section{Score Matching Theory}
\label{sec:app:score_matching_theory}

\subsection{Preliminary mathematical results}
\subsubsection{Bias-variance decomposition}
Let $p(\xv)$ be a probability density function on a finite dimensional vector space $\mathcal{X}$ with Euclidean norm denoted $\|\cdot\|$. Define $\bm{\mu} = \int \xv p(\xv) \, d\xv$.
\paragraph{Claim}
For any $\sv \in \mathcal{X}$,
\begin{equation}
     \int p(\xv) \| \sv - \xv \|^2 \,d\xv = \| \sv - \bm{\mu}\|^2 +  \int p(\xv) \| \bm{\mu} - \xv \|^2 \,d\xv.
\end{equation}

\begin{proof}
First, we have
\begin{align}
\| \sv - \xv \|^2 &= \| \sv -\bm{\mu}\|^2 + 2 (\sv - \bm{\mu}) \cdot (\bm{\mu} - \xv)  + \| \bm{\mu}- \xv \|^2.
\end{align}
Then,
\begin{align}
\int p(\xv) (\sv - \bm{\mu}) \cdot (\bm{\mu} - \xv) \,d\xv &= (\sv - \bm{\mu}) \cdot \int p(\xv) (\bm{\mu} - \xv) \,d\xv \\
&=(\sv - \bm{\mu}) \cdot \left( \bm{\mu} - \int \xv p(\xv) \, d\xv \right) \\
&=0.
\end{align}
This completes the proof.
\end{proof}

\paragraph{Corollary}
The mean value minimises the mean squared error
\begin{equation}
     \bm{\mu} = \min_\sv \int p(\xv) \| \sv - \xv \|^2 \,d\xv.
\end{equation}

\subsubsection{Stein's Identity}
Let $p(\xv_2,\dots,\xv_N \mid \phi)$ be any family of probability densities parametrized by $\phi$.
\paragraph{Claim}
\begin{equation}
     \int  p(\xv_2 \dots, \xv_N \mid \phi) \, \nabla_\phi  \log p(\xv_2 \dots, \xv_N \mid \phi)  \, d \xv_{2:N} =0.
\end{equation}

\begin{proof}
\begin{align}
0 &= \nabla_{\phi} \int   p(\xv_2 \dots, \xv_N \mid \phi) \, d \xv_{2:N} \\
&=  \int \nabla_{\phi}  p(\xv_2 \dots, \xv_N \mid \phi) \, d \xv_{2:N} \\
&=  \int  p(\xv_2 \dots, \xv_N \mid \phi) \, \nabla_\phi  \log p(\xv_2 \dots, \xv_N \mid \phi)  \, d \xv_{2:N}.
\end{align}
\end{proof}

\subsection{Main result}
We write $p(\xv_1,\dots, \xv_N)= |\Psi(\xv_1,\dots,\xv_N)|^2$ for the joint probability density of the $N$ electrons. Then the electron density
\begin{equation}
    \rho(\xv) = N\int p(\xv,\xv_2,\dots,\xv_N)\,d\xv_{2:N}
\end{equation}
is the marginal probability density multiplied by $N$.
Also recall that $\xv = (\rv, \sigma)$ where $\rv\in\mathbb{R}$ is the spatial co-ordinate and $\sigma \in \{\uparrow,\downarrow\}$ is the spin.
The \emph{score function} is $\nabla_\rv \log \rho(\xv)$, the gradient of the density with respect to position.

We now prove that Eq. (13) is a correct loss function to learn $\rho$ in two claims.

\paragraph{Claim}
\begin{equation}
    \label{eq:main-result}
    \nabla_\rv \log \rho = \argmin_{s: \mathbb{R}^3\times \{\uparrow,\downarrow\}\to\mathbb{R}^3}  \int \left\| \nabla_{\rv_1} \log p(\xv_1, \dots, \xv_N) - s(\xv_1) \right\|^2 p(\xv_1, \dots, \xv_N) \,d \xv_{1:N} .
\end{equation}
where the minimization is over functions $s$.

\begin{proof}
For a given $\xv_1$, the minimization problem is solvable and the minimizing value is the conditional expectation. This follows from the corollary of the bias-variance decomposition. Then,
\begin{align}
    \label{eq:fisher-identity-proof-begin}
    s(\xv_1) &= \int  p(\xv_2 \dots, \xv_N \mid \xv_1) \, \nabla_{\rv_1} \log p(\xv_1, \dots, \xv_N)  \, d \xv_{2:N} \\
    &= \int  p(\xv_2 \dots, \xv_N \mid \xv_1) \, \nabla_{\rv_1} \left( \log \rho(\xv_1) + \log p(\xv_2 \dots, \xv_N\mid \xv_1) \right)  \, d \xv_{2:N} \\
    &= \nabla_{\rv_1} \log \rho(\xv_1) +  \int  p(\xv_2 \dots, \xv_N \mid \xv_1) \, \nabla_{\rv_1}  \log p(\xv_2 \dots, \xv_N \mid \xv_1)  \, d \xv_{2:N} \\
    &= \nabla_{\rv_1} \log \rho(\xv_1), \label{eq:fisher-identity-proof-end}
\end{align}
using Stein's Identity in the last step (taking $\xv_1$ as our $\phi$).
\end{proof}

\paragraph{Corollary}
We now use the fact that $p(\xv_1,\dots,\xv_N)$ is exchangeable, i.e.~for any permutation $\sigma$, $p(\xv_1,\dots,\xv_N) = p(\xv_{\sigma(1)},\dots,\xv_{\sigma(N)})$. This follows from the antisymmetry of the wave function.
Then
\begin{align}
&\int \frac{1}{N}\sum_{i=1}^N \left(\left\| \nabla_{\rv_i} \log p(\xv_1, \dots, \xv_n) - s(\xv_i) \right\|^2 \right) p(\xv_1, \dots, \xv_n) \,d \xv_{1:N} \\
&=\frac{1}{N}\sum_{i=1}^N \int  \left\| \nabla_{\rv_i} \log p(\xv_1, \dots, \xv_n) - s(\xv_i) \right\|^2 p(\xv_1, \dots, \xv_n) \,d \xv_{1:N} \\
&=\frac{1}{N}\sum_{i=1}^N \int  \left\| \nabla_{\rv_i} \log p(\xv_i, \xv_{\setminus i}) - s(\xv_i) \right\|^2 p(\xv_i, \xv_{\setminus i}) \, d \xv_{1:N} \\
&= \int  \left\| \nabla_{\rv_1} \log p(\xv_1, \xv_{2:N}) - s(\xv_1) \right\|^2 p(\xv_1, \xv_{2:N}) \, d \xv_{1:N}.
\end{align}

\subsubsection{Unbiasedness of the training gradient estimator}
Now assume $s(\cdot; \bm \theta)$ is parameterized and we aim to find the minimizer of \eqref{eq:main-result} via stochastic gradient descent. Denote
\begin{align}
    \ell(\bm \theta) &= \mathbb{E}_{p(\xv_1, \xv_{2:N})} \left[ \| \nabla_{\rv_1} \log p(\xv_1, \xv_{2:N}) - s(\xv_1; \bm \theta) \|^2 \right] \\
    \hat \ell(\bm \theta) &= \mathbb{E}_{p(\xv_1)}\left[\| \nabla_{\rv_1} \log p(\xv_1) - s(\xv_1; \bm \theta) \|^2 \right].
\end{align}
\paragraph{Claim}
\begin{align}
    \nabla_{\bm \theta} \ell(\bm \theta) = \nabla_{\bm \theta} \hat \ell(\bm \theta)
\end{align}

\begin{proof}
Analyzing the residual gives
\begin{align}
    \ell(\bm \theta) - \hat \ell(\bm \theta)
    &= \mathbb{E}_{p(\xv_1, \xv_{2:N})}\left[ \| \nabla_{\rv_{1}} \log p(\xv_1, \xv_{2:N}) \|^2 - \| \nabla_{\xv_{1}} \log p(\xv_1) \|^2 - 2 \bm \delta(\xv_1, \xv_{2:N})^\top s(\xv_1; \bm \theta)\right]
\end{align}
with 
\begin{align}
    \bm \delta(\xv_1, \xv_{2:N}) := \nabla_{\rv_1} \log p(\xv_1, \xv_{2:N}) - \nabla_{\rv_1} \log p(\xv_1).
\end{align}
Using eqs. \eqref{eq:fisher-identity-proof-begin} - \eqref{eq:fisher-identity-proof-end} we get
\begin{align}
    \mathbb{E}_{p(\xv_1, \xv_{2:N})}\left[ \bm \delta(\xv_1, \xv_{2:N}) \right] &= \bm 0
\end{align}
and as such
\begin{align}
    \nabla_{\bm\theta} \left(\ell(\bm \theta) - \hat \ell(\bm \theta)\right) = \nabla_{\bm\theta}  \mathbb{E}_{p(\xv_1, \xv_{2:N})}\left[ \| \nabla_{\xv_{1}} \log p(\xv_1, \xv_{2:N}) \|_2^2 - \| \nabla_{\xv_{1}} \log p(\xv_1) \|_2^2 \right] = \bm 0.
\end{align}
\end{proof}

\subsection{Connection to force matching}
\label{sec:app:force_matching}

We note a strong connection between our approach to electron density estimation and the force matching approach to coarse-graining \citep{noid2008multiscale}.
Force matching considers a more general coarse-graining operator that is a linear map $\pi$ from the original co-ordinates $\xv_1,\dots,\xv_N$ of a system to a smaller set of co-ordinates that describe certain aspects of the system.
Our approach can be seen as a special case of coarse-graining where we use the operator
\begin{equation}
    \xv_1,\dots,\xv_N \mapsto \xv_1.
\end{equation}

\section{Kato's cusp condition}
\label{sec:app:cusp}

We use the shorthand $\mathbf{r}_I = \mathbf{r} - \Rv_I$ and $r_I = \|\mathbf{r}_I\|$. The cusp term in the model is given by
\begin{align}
    C(\mathbf{r}) &= \sum_I 2\sqrt{\pi} \text{erf}(\tfrac{1}{2} Z_I r_I) =\sum_I  4 \int_0^{\tfrac{1}{2} Z_I r_I} e^{-t^2} dt.\\
    \intertext{The gradient of this term is then}
    \nabla_{\mathbf{r}} C(\mathbf{r}) &= \sum_I 2Z_I e^{-\tfrac{1}{4} Z_I^2 r_I^2}  \hat{\mathbf{r}_I}. \\
    \intertext{The Laplacian can be computed using standard formulae for spherical co-ordinates}
    \nabla^2_{\mathbf{r}} C(\mathbf{r}) &= \sum_I \frac{1}{r_I^2}\frac{\partial}{\partial r_I}\left( r_I^2 \frac{\partial C}{\partial r_I} \right) \\
    &= \sum_I \frac{1}{r_I^2}\frac{\partial}{\partial r_I}\left( 2Z_I r_I^2 e^{-\tfrac{1}{4} Z_I^2 r_I^2} \right) \\
    &= \sum_I \frac{1}{r_I^2}\left( 4Z_I r_I - Z_I^3 r_I^3  \right)e^{-\tfrac{1}{4} Z_I^2 r_I^2} \\
    &= \sum_I \left( \frac{4Z_I}{r_I} - Z_I^3 r_I  \right)e^{-\tfrac{1}{4} Z_I^2 r_I^2}, \label{eq:useful_laplacian} \\
    \intertext{furthermore, the asymptotic behaviour for small $r_I$ is}
    &= \sum_I \left( \frac{4Z_I}{r_I} - Z_I^3 r_I - Z_I^3 r_I + O(r_I^3)  \right)\\
    &= \sum_I \left( \frac{4Z_I}{r_I} - 2Z_I^3 r_I + O(r_I^3)  \right).
    \label{eq:lap_cusp}
\end{align}

\section{Effective potential computation}
\label{sec:app:v_eff}
For the LiH experiment in Sec. IVB, we compute the following quantity 
\begin{equation}
    v_\mathrm{eff}(\xv) - v_\mathrm{ext}(\xv) = \frac{1}{8} \|\nabla_\rv \log \rho(\xv)\|^2 + \frac{1}{4} \nabla^2_\rv \log \rho(\xv) + \sum_I \frac{Z_I}{r_I}.
\end{equation}
Using the additive form of our density $U_\phi(\xv)=E_\phi(\xv)+M_\phi(\xv)+C(\rv)$, gives
\begin{align}
    v_\mathrm{eff}(\xv) - v_\mathrm{ext}(\xv) = \frac{1}{8}\|-\nabla_\rv U_\phi(\xv)\|^2 - \frac{1}{4}\nabla_\rv^2 \left(E_\phi(\xv)+M_\phi(\xv)+C(\rv) \right)+\sum_I \frac{Z_I}{r_I} .
\end{align}
From this equation, we use the fact that $\|-\nabla_\rv U_\phi\|^2, \nabla_\rv^2 E_\phi$, and $\nabla_\rv^2 M_\phi$ are bounded by design. Cancellation of the poles in the external potential can therefore only come from $\nabla_\rv^2 C$. We expand this Laplacian using the analysis from above, Eq.~\eqref{eq:useful_laplacian}, to give
\begin{align}
\begin{split}
\label{eq:veff_m_vext}
    v_\mathrm{eff}[\rho](\xv) - v_\mathrm{ext}[\rho](\xv) =& \ \frac{1}{8}\|-\nabla_\rv U_\phi(\xv)\|^2 - \frac{1}{4}\nabla_\rv^2 \left(E_\phi(\xv)+M_\phi(\xv) \right)  \\ &+ \sum_I \left( \frac{-Z_I}{r_I} (e^{-\frac{1}{4}Z_I^2 r_I^2} - 1) + \tfrac{1}{4}Z_I^3r_I e^{-\frac{1}{4}Z_I^2 r_I^2}\right).
\end{split}
\end{align}
Note that
\begin{equation}
\label{eq:veff_m_vext2}
    \frac{-Z_I}{r_I} (e^{-\frac{1}{4}Z_I^2 r_I^2} - 1) = \frac{1}{4} Z_I^3 r_I + O(r_I^3)
\end{equation}
no longer explodes as $r_I \to 0$.
For numerical stability, we explicitly replace $\frac{-Z_I}{r_I} (e^{-\frac{1}{4}Z_I^2 r_I^2} - 1)$ with the first term of its series expansion for small $r_I$.

\section{Total variation distance}
The total variation (TV) distance is a metric on the space of probability distributions \citep{tsybakov2009} that can be expressed as
\begin{equation}\label{eq:TV_distance}
    \mathrm{TV}(\rho,\rho') = \frac{1}{2} \int \abs{\rho(\rv) - \rho'(\rv)} d\rv.
\end{equation}
We extend this definition to electron densities (keeping their normalization to $N$, rather than 1).
Whilst it is generally not possible to find ``ground truth'' electron densities, the TV distance is helpful to compare to references and to check expected symmetries of the density.

\section{Estimation of observables from electron density}
\label{sec:app:observables}

We can also validate our \methodabb~models by estimating certain one-body observable quantities.
In every case, the observable quantity involves an expectation over the density, $\int \cdot \ \rho(\xv) \, d\xv$.
We compute these integrals using Lebedev--Laikov grids \citep{lebedev1999quadrature}.

By the Hellmann-Feynman Theorem \citep{feynman1939forces,politzer2018hellmann}, the \emph{force} on nucleus $I$ is given by
\begin{equation}
\label{eq:hellmann-feynman}
    \mathbf{F}_I=\sum_{J \ne I} \frac{Z_IZ_J (\Rv_I - \Rv_J)}{\|\Rv_I - \Rv_J\|^3} + \int  \frac{Z_I(\rv - \Rv_I)}{\|\rv - \Rv_I\|^3}\rho(\xv) \, d\xv\,.
\end{equation}
This formula holds when $\rho$ corresponds to a variationally optimal wave function.
We compare the approach of integrating $\rho$ obtained from our \methodabb~model with the estimators 
that directly operate on samples from the deep QMC wave function.

Second, we can compute the expectation value of the \emph{nuclear potential}
\begin{equation}\label{eq:nuclear_potential}
    V[\rho] = -\int \left(\sum_I \frac{Z_I }{\norm{\rv-\Rv_I}}\right)\rho(\xv) \,d\xv;
\end{equation}
this is the only component of the Hamiltonian that can be directly calculated from the density, and is most sensitive to density quality near nuclei.

Third, \emph{quadrupole moment}, $Q$, are defined as
\begin{align}\label{eq:moments}
    Q_{ij} = \int \left(3r_{i}r_{j} - \|\rv\|^2 \delta_{ij} \right) \rho(\xv)\,d\xv,
\end{align}
and help describe the electrostatic properties of a charge density.

\section{Classical VMC density estimators}\label{sec:app:density_vmc_estimator}
In this section, we introduce the equations to evaluate two classical VMC density estimators mentioned in the main text, i.e., Assaraf--Caffarel--Scemama zero bias estimator (ACS-ZB) and Per--Snook--Russo zero-variance and approximate zero-bias estimator (PSR-ZV(ZB)), at any point $\rv^\prime$ in 3D space. ACS-ZB density estimator \cite{assaraf_improved_2007} is unbiased for arbitrary $f(\rv_i=\rv^\prime)=1$ and defined by 
\begin{equation}
\rho(\rv^\prime)=\left\langle \sum_{i}-\frac{1}{4\pi}\frac{1}{||\rv^{\prime}-\rv_{i}||}\frac{\nabla_{i}^{2}(f\Psi^{2})}{\Psi^{2}}\right\rangle _{\rv\sim\Psi^{2}}\
\end{equation}

To reduce the variances of this estimator for certain $\rv^\prime$, Håkansson and Mella derived the auxiliary function $f$ by using known physical properties for two specific regions as
\begin{equation}
f=1+2Z_{I}(||\rv_{i}-\Rv_I||-||\rv^{\prime}-\Rv_I||)
\end{equation}
for density near nuclear $\Rv_I$ ($||\rv^{\prime}-\Rv_I||\rightarrow0$), 
and 
\begin{equation}
f=(1+\lambda||\rv^{\prime}-\rv_{i}||)e^{-2\kappa_{\sigma}||\rv^{\prime}-\rv_{i}||}
\end{equation}
for $||\rv^{\prime}-\Rv_I||\rightarrow\infty$. 

PSR-ZV(ZB) \cite{per2011zero} for VMC is defined as follows:
\begin{equation}
    \rho(\rv^{\prime})=\left\langle \sum_{i}-\frac{1}{4\pi}\nabla_{i}^{2}\frac{1}{||\rv^{\prime}-\rv_{i}||}-\frac{1}{2}(\nabla_{i}^{2}f+\nabla_{i}f\nabla_{i}\log\Psi^{2})+2f(E_\text{loc} - E)\right\rangle _{\rv\sim\Psi^{2}}, 
\end{equation}
where $E_\text{loc}$ is the local energy of the wave function and $E$ is the energy expectation value of the wave function. To make this estimator density zero-variance and approximate zero-bias at nucleus $\Rv_I$,
$f$ is chosen as 
\begin{equation}
f=-\frac{1}{2\pi}\sum_{i}\frac{1}{||\rv_{i}-\Rv_I||}+\frac{Z_{I}}{\pi}\sum_{i}\log(||\rv_{i}-\Rv_I||)+\frac{Z_{I}^{2}}{\pi}\sum_{i}||\rv_{i}-\Rv_I||.
\end{equation}

\section{Assaraf--Caffarel force estimator for the wave function}\label{sec:app:ac_zv}
Forces in quantum Monte Carlo can be evaluated by applying the Hellmann-Feynman theorem, turning the energy derivative into an expectation value of the ``classical" Coulomb force over samples from the wave function
\begin{equation}
    \mathbf{F}_I = \sum_{J \ne I} \frac{Z_IZ_J (\Rv_I - \Rv_J)}{\|\Rv_I - \Rv_J\|^3} + \Big\langle  \frac{Z_I(\rv - \Rv_I)}{\|\rv - \Rv_I\|^3}\Big\rangle_{\rv \sim \Psi^2}\,.
\end{equation}
This estimator of the Hellman-Feynman force has the desired expectation value, but exhibits infinite variance due to the divergences of the force at the nuclei, severely limiting its practical applicability. To improve the efficiency, Assaraf and Caffarel proposed a force estimator that has the same mean but finite variance
\begin{equation}
    \mathbf{F}^{ZV}_{I} =  \mathbf{F}_I + \Big\langle\frac{(H-E_\text{loc})\Tilde{\Psi}}{\Psi}\Big\rangle_{\rv \sim \Psi^2} \,,
\end{equation}
where 
$ \Tilde{\Psi}$ is an approximation of the wave function derivative \cite{assaraf2003zero}.
The ``minimal" version of the zero variance estimator cancels the divergences in the Coulomb force by choosing
\begin{equation}
     \Tilde{\Psi}_\text{min} = Q\Psi~~~\text{with}~~~Q_{I,j}=Z_I\sum_{i}\frac{(\rv_{i}-\Rv_I)_j}{\|\rv_{i}-\Rv_I\|}\,,
\end{equation}
where $I$ indexes atoms and $j$ indexes the three spacial dimensions.
Within this approximation the expectation value takes the form
\begin{equation}
     \mathbf{F}^{ZV}_{I} = \Big\langle -\nabla_{\mathbf r} Q\frac{\nabla_{\mathbf r} \Psi}{\Psi}\Big\rangle_{\rv \sim \Psi^2}\,.
\end{equation}
The estimator can be further extended with a term to correct a potential bias in the force estimate due to a discrepancy of the wave function from the true ground state
\begin{equation}
    \mathbf{F}^{ZVZB}_{I} = \mathbf{F}^{ZV}_{I} + \Big\langle2(E_\text{loc} - E)\frac{\Tilde\Psi}{\Psi}\Big\rangle_{\rv \sim \Psi^2}\,.
\end{equation}
Note that the $\mathbf{F}^{ZVZB}_{I}$ estimator requires the explicit evaluation of local energies, adding a significant computational overhead over the $\mathbf{F}^{ZV}_{I}$ estimator (see\phantom{-}\citet{qian2022interatomic} for more details). 

\section{Additional experiments}
\label{sec:app:atom}
\subsection{Kato's cusp condition using $v_\text{eff}-v_\text{ext}$ vs $r$}
\begin{figure}[ht]
    \centering
    \includegraphics[width=0.5\columnwidth]{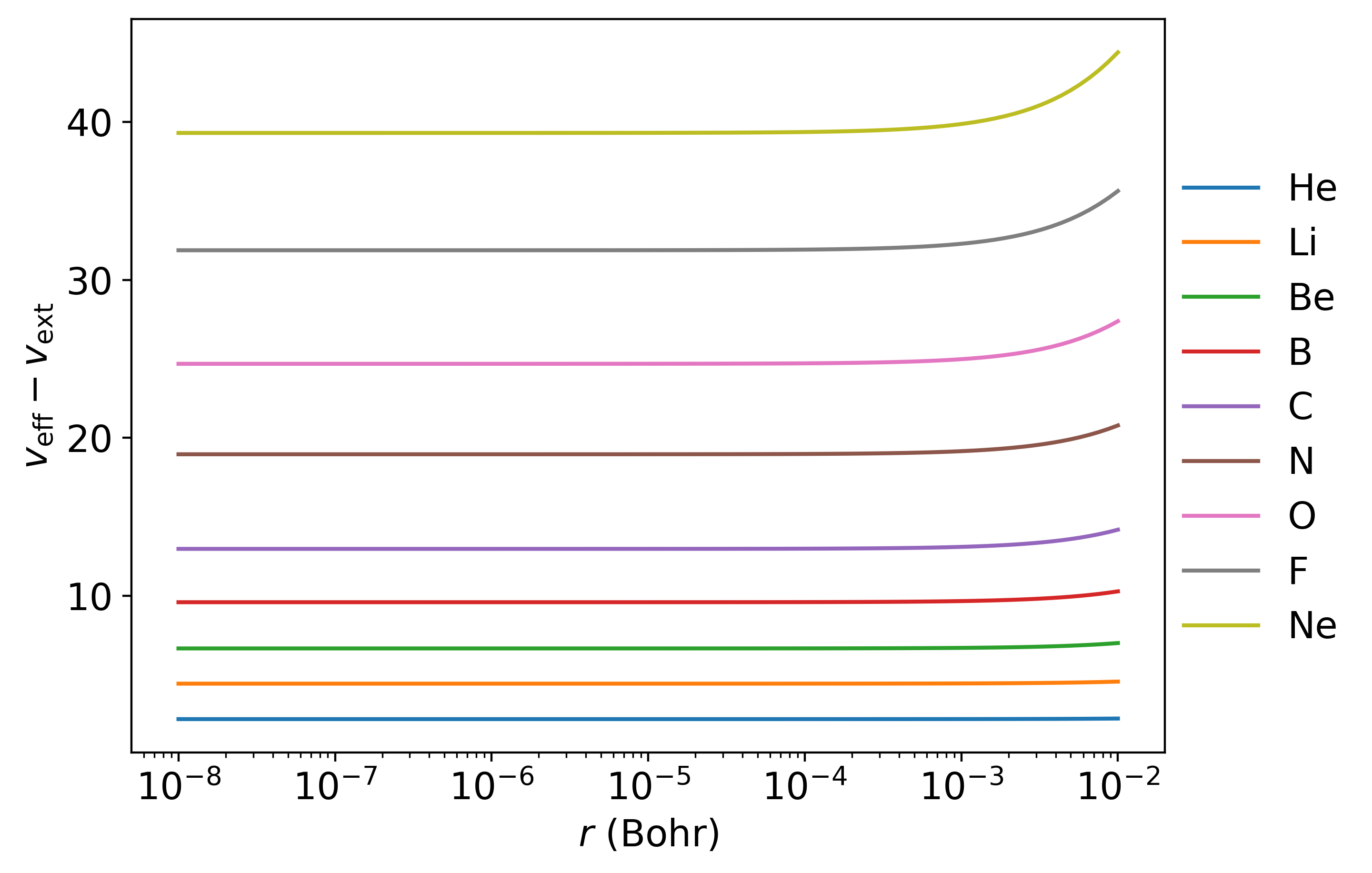}
    \caption{$v_\text{eff}-v_\text{ext}$ vs $r$ for 9 small atoms examined in this study. The x-axis is plotted on a log-scale to visualize the distances to nuclei}
    \label{fig:atom_kato}
\end{figure}

In Fig. \ref{fig:atom_kato}, the $v_\text{eff}-v_\text{ext}$ values are plotted as a function of $r=|\rv-\mathbf{R}_I|$ for the small atoms tested in this study. For all atoms, $v_\text{eff}-v_\text{ext}$ values approach constants when $r \to 0$ as expected (Eq. ~\ref{eq:veff_m_vext} and \ref{eq:veff_m_vext2}). This further verifies that the neural network densities have the correct behavior near the nucleus and satisfy an exact Kato's cusp condition.

\begin{figure}[ht]
    \centering
    \includegraphics[width=0.5\columnwidth]{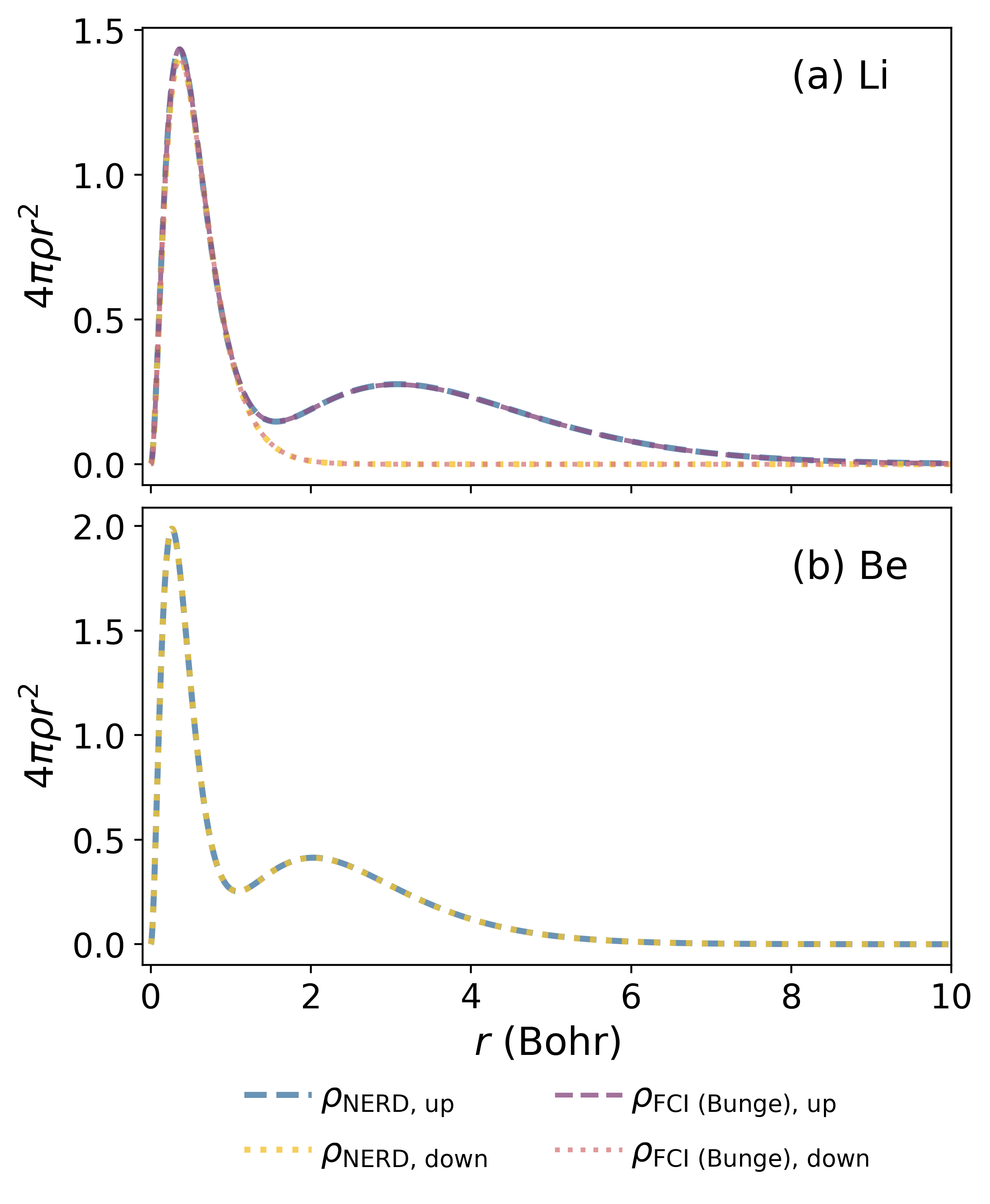}
    \caption{Comparison of the $\rho_\text{up}$ and $\rho_\text{down}$ of (a) \ch{Li} and (b) \ch{Be}. To better visualize the differences between spin-up and spin-down densities, all the values are plotted as $4\pi\rho r^2$. The FCI spin density values for Li are computed with 8s up to k basis functions, and are kindly provided by Dr. Carlos Bunge.}
    \label{fig:spin_densities}
\end{figure}

\subsection{Spin-up and spin-down density comparisons for \ch{Li} and \ch{Be}}
We note that \methodabb~models could provide accurate total densities and also spin densities. Figure \ref{fig:spin_densities} displays comparisons of spin {\methodabb}s for \ch{Li} and \ch{Be} using $4\pi\rho r^2$ values, where $r=|\rv-\mathbf{R}_I|$ the corresponding FCI spin densities for Li are also compared.

\subsection{Quadrupole moment evaluations for benzene}
\begin{figure}[h]
    \centering
    \includegraphics[width=0.4\columnwidth]{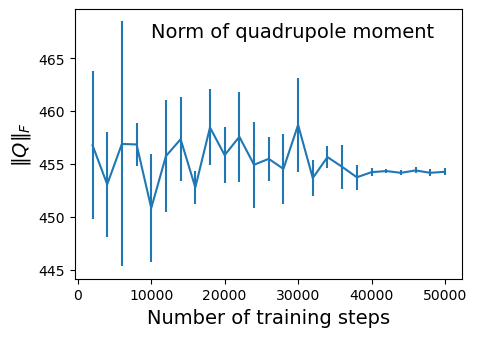}
    \caption{Norm of the quadrupole moment for equilibrium benzene structure}
    \label{fig:quadrupole_moment}
\end{figure}

Figure \ref{fig:quadrupole_moment} plots an additional experiment (similar to Fig. 9) on the convergence of quadrupole moments with training steps for benzene.

\section{Quantum chemistry calculations}
\label{sec:app:qc}

All the densities from second quantized quantum chemistry methods in this study are obtained by PySCF 2.4.0 \cite{sun2018pyscf} with the default settings. The calculations include FCI/aug-cc-pVQZ densities of Li and Be atoms (Figs. 2 and 3), FCI/aug-cc-pVTZ densities of the equilibrium structure of \ch{LiH} with a bond length of 3.015 Bohr (Fig. 5), and CCSD/aug-cc-pVQZ densities of small atoms in Table 2 and \ch{H4} square with bond lengths of 3.0236 Bohr (Fig. 6). The literature VMC densities of Be and Ne atoms are obtained from\phantom{-}\citet{filippi1996recent}

\section{Experimental settings for Psiformer and \methodabb~model training}
\label{sec:app:exp_setting}

The following tables summarize the experimental settings of the Psiformer (Table~\ref{tab:psiformer_settings}) and the \methodabb~models (Table~\ref{tab:density_settings}) reported in this study.
For all experiments, optimization during variational wave function training was done using KFAC \citep{martens2015optimizing} with learning rate $5\times 10^{-2}$ with geometric decay schedule, damping factor $10^{-3}$, and norm constraint $10^{-3}$. Supervised pretraining to Hartree-Fock used the Adam optimizer with learning rate $10^{-3}$.
Density training also used the Adam optimizer.

\begin{table}[htbp]
\centering
\caption{Experimental settings of Psifomer models for different systems.}\label{tab:psiformer_settings}
\begin{tabular}{lllllll}  
\hline  
 Setting &Atoms \& ions &LiH dissociation &\ch{H4} &Benzene & \ch{H2} & \ch{Li2} \& \ch{N2} \\  
\hline
Electron batch size &4096 &2048  &4096 & 4096 & 2048 & 4096 \\  
MCMC sampler &MALA\textsuperscript{\emph a} & MALA & MALA & MHA \textsuperscript{\emph b} & MHA &  MALA \\ 

Pretraining steps &20000 &20000 &500 & 20000 & 5000 & 5000 \\  

Training steps & max(50k, 10k$\, \cdot \,n_{e^-}$) & 100k & 100k & 200k & 100k & 200k \\   
Energy evaluation Steps & 10k  & ---  & --- & --- & 10k & 10k \\
Force evaluation Steps & ---  & ---  & --- & --- &  10k & 10k \\
Dipole moment evaluation Steps & ---  & 10k  & --- & --- &  --- & --- \\
\hline
\multicolumn{6}{l}{\textsuperscript{\emph a} Metropolis adjusted Langevin algorithm. \textsuperscript{\emph b} Metropolis--Hastings algorithm.}
\end{tabular}  
\end{table}  

\begin{table}[htbp] 
\centering
\caption{Experimental settings of \methodabb~models for different systems.}\label{tab:density_settings}
\begin{tabular}{llllll}  
\hline  
Setting &Atoms &LiH dissociation &\ch{H4} &Benzene & \ch{H2}, \ch{Li2} \& \ch{N2} \\ 
\hline  
Electron batch size &4096 &2048  &2048 & 4096 & 4096 \\ 
MCMC sampler &MHA &MHA  &MHA & MHA & MHA \\ 
Steps & max(25k, 5k$\, \cdot \, n_{e-}$) & 25k & 25k & 50k & 40k \\
NCE weight, $\lambda$ & 1 &  1 & 1 & 1  & 1 \\
Learning rate & \multicolumn{5}{c}{Cosine decay schedule: init value=0.01, decay steps=0.8 $\cdot$ steps, $\alpha$=6e-5, b1=0.95, and b2=0.999} \\
\hline  
\end{tabular}  
\end{table}
\section{Computational cost of \methodabb~model training}
In general, the computational effort for obtaining NERD densities can be split into the cost of calculating the reference VMC solution, and the density estimation, with the former typically being the predominant factor. 
The NERD density estimation can then itself be split into the process of obtaining electron samples and gradients from the reference wave function, and the actual optimization of the density model using these samples. 
Again, sampling the wave function and computing its gradients is the more time consuming part, contributing 98\% to the computational cost of the density estimation in our setup.
Note that while the cost of evaluating gradients increases with the number of electrons, more electrons result in a linearly growing number of one-electron samples for the density estimation for every gradient evaluation.
To exemplify the cost of the density estimation we list the time per iteration on a single Nvidia-A100 GPU using a batch size of 4096 in Table~\ref{tab:cost}.
Combining these examples with the training setting listed in Table~\ref{tab:density_settings} the total cost of the density optimization can be estimated.
\begin{table}[htbp] 
\centering
\caption{Iteration speed of density estimation on Nvidia-A100 GPU.}\label{tab:cost}
\begin{tabular}{lrr}  
\hline  
System & Time\\
\hline  
H2 & 0.241 s/it \\
Li2 & 0.396 s/it \\
N2 & 0.769 s/it \\
\hline  
\end{tabular}  
\end{table}
\clearpage
\bibliography{biblio}